\documentclass[aps,pra,twocolumn,groupedaddress]{revtex4}
\usepackage{epsfig}
\usepackage{graphicx}
\usepackage{amsfonts}
\usepackage{latexsym}
\usepackage{bm}

\def\8{\infty}
\def\ohm{\frac{1}{2m}}
\def\oh{\frac{1}{2}}
\def\ot{\frac{1}{3}}

\def\d{\partial}

\def\undertext#1{\vtop{\hbox{#1}\kern 1pt \hrule}}

\def\dbyd#1#2{\frac{d#1}{d#2}}

\def\pbyp#1#2{\frac{\partial#1}{\partial#2}}

\def\be{\begin{equation}}
\def\ee{\end{equation}}
\def\bea{\begin{eqnarray} & &}
\def\eea{\end{eqnarray}}

\def\rf#1{(\ref{#1})}

\def\rf#1{(\ref{#1})}
\def\t{\tilde}

\def\sign{{\rm sign}}

\def\rfs#1{Eq.~\rf{#1}}

\begin{document}


\title{One Dimensional Gas of Bosons with Feshbach Resonant
Interactions}


\author{V. Gurarie}
\affiliation{Department of Physics, CB390, University of Colorado,
Boulder CO 80309}


\date{\today}

\begin{abstract}
We present a study of a gas of bosons confined in one dimension with
Feshbach resonant interactions, at zero temperature. Unlike the gas
of one dimensional bosons with non-resonant interactions, which is
known to be equivalent to a system of interacting spinless fermions
and can be described using the Luttinger liquid formalism, the
resonant gas possesses novel features. Depending on its parameters,
the gas can be in one of three possible regimes. In the simplest of
those, it can still be described by the Luttinger liquid theory, but
its Fermi momentum cannot be larger than a certain cutoff momentum
dependent on the details of the interactions. In the other two
regimes,  it is equivalent to a Luttinger liquid at low density
only. At higher densities its excitation spectrum develops a
minimum, similar to the roton minimum in helium, at momenta where
the excitations are in resonance with the Fermi sea. As the density
of the gas is increased further, the minimum dips below the Fermi
energy, thus making the ground state unstable. At this point the
standard ground state gets replaced by a more complicated one, where
not only the states with momentum below the Fermi points, but also
the ones with momentum close to that minimum, get filled, and the
excitation spectrum develops several branches. We are unable so far
to study this new regime in detail due to the lack of the
appropriate formalism.
\end{abstract}


\maketitle


\section{Introduction}
A gas of one dimensional interacting bosons is an interesting system
to study, and we have in our possession a number of techniques to
study it exactly. One of the first papers to study this system,
Ref.~\cite{Girardeau1960}, demonstrated that with infinite repulsive
interactions, it is equivalent to a gas of noninteracting spinless
fermions. A subsequent paper Ref.~\cite{Lieb1963} showed that even
if the interactions are not infinite, as long as they can be
approximated by short ranged delta function type interactions, the
gas of interacting bosons can still be studied with the help of a
method closely related to Bethe Ansatz. Its resulting behavior is in
many ways equivalent to a gas of interacting fermions. In
particular, two particles cannot have the same momentum when far
away from each other, which is a sort of a Pauli principle for
bosons. In the ground state all the momentum states below the Fermi
momentum are occupied, while the rest are empty.

Subsequently it was demonstrated in Ref.~\cite{Haldane1981} that the
excitations of this gas around the Fermi points could be described
by the effective ``bosonized" theory which was later  named
Luttinger liquid theory. It is applicable to both systems of
interacting bosons and fermions in one dimension.

Interest in such systems has been revived recently due to the
emergence of new experimental techniques where the one dimensional
gas of interacting bosons can be manufactured in a laboratory with
ultracold bosonic atoms as its constituent particles
\cite{Paredes2004}. For the first time it has become possible to
study these gases experimentally.

A typical interaction between ultracold atoms is due to van der
Waals forces, which are relatively short ranged compared with
typical interatomic separation in a cold atom experiment, thus it is
not impossible to conjecture that Bethe Ansatz techniques may be
applicable to study the behavior of a gas of such atoms in one
dimension. It has also become possible to control the interactions
with the help of Feshbach resonances \cite{Timmermans1999}. Feshbach
resonance allows to bring a pair of atoms in a resonance with a
diatomic molecule, whose binding energy can be controlled by the
application of magnetic field. The binding energy can be both
negative, in which case the atoms will want to release their energy
and form a stable molecule, or positive, in which case the molecular
lifetime is finite and the molecule eventually decays into a pair of
atoms.

It is well known that, in one dimension, if the molecule's energy is
negative, and if the atoms are bosons, then the system is unstable
\cite{Lieb1963} and collapses into a state whose energy is negative
and  is proportional to the square of the number of particles. Thus
the energy per particle is essentially minus infinity in the
thermodynamic limit. This situation cannot be experimentally
realized.

If, however, the energy of the molecule is positive, then this
results in the new type of atomic gas not studied in the literature
before. The study of the gas with resonant interactions is the
subject of this paper.

The Hamiltonian of such a gas can be written in the following form
\cite{Timmermans1999}
\begin{eqnarray} \label{eq:FA}
H &= & \int dx~\left[\frac{1}{2m_a}  \, \partial_x a^\dagger \, \d_x
a+\frac{1}{4m_a} \, \d_x b^\dagger \, \d_x b + \epsilon_0 b^\dagger
b+ \right. \cr && \left. + \frac{c}{2} a^\dagger a^\dagger a a +
\frac{g}{\sqrt{2}} \left( b \, {a^\dagger}^2+b^\dagger a^2
\right)\right],
\end{eqnarray}
where $a^\dagger$, $b^\dagger$,  and $a$, $b$ are bosonic atomic and
molecular creation and annihilation operators. Here the
interactions, given by the Feshbach term linear in $b$ and quadratic
in $a$, are automatically short ranged. The mass of atoms $a$ is
$m_a$, while the molecules $b$, being able to covert into two
$a$-particles, must twice as large mass. Otherwise, \rfs{eq:FA}
would violate Galilean invariance. $\epsilon_0$ is the parameter
which controls the binding energy of the molecules.

We note that the Hamiltonian \rfs{eq:FA} includes both the Feshbach
$g$ term and the delta-function interaction $c$ term. Although often
added when studying Feshbach resonant gas in higher dimensions, the
$c$-term can be shown to be unnecessary  in three dimensional space
\cite{Gurarie2005}. It can be absorbed into the redefinition of
$\epsilon_0$ and $g$. However, in one dimension the situation is
different, and the delta-function term $c$ is crucial.  If the
 $c$ term is omitted in \rfs{eq:FA}, then there exists
a real bound state of atoms with negative binding energy at all
values of $\epsilon_0$ and $g$. This particular situation  does not
arise in three dimensions. It is related to the fact that in one
dimensional space (as well as in two dimensions) there exists a
bound state in an arbitrary weak attractive potential. A bosonic gas
with real bound state of atoms is unstable, as was discussed above.

Fortunately real atoms always repel each other at short distances,
and this is encoded by the $c$-term in \rfs{eq:FA}. This term is
necessary to create a situation where real bound states are absent,
while resonances, molecules with positive binding energy and finite
lifetime in resonance with atomic gas, are present.

A version of the Hamiltonian \rfs{eq:FA} in higher than one
dimensional space was analyzed in Ref.~\cite{Radzihovsky2004}. It is
well known though that the one dimensional physics is usually quite
different from its multidimensional counterpart.  In one dimensional
space a similar Hamiltonian was studied in Ref.~\cite{Sheehy2005}.
However, in that paper the $a$-particles were spin-1/2 fermions, and
the analysis presented  in that paper was based on the bosonization
techniques, which do not capture the excitation spectrum beyond the
vicinity of the Fermi points. In this paper we study \rfs{eq:FA}
exactly, and our techniques do not rely on bosonization. So we are
able to obtain the exact excitation spectrum of \rfs{eq:FA}.

The question of whether the techniques presented here are also
suitable to study the gas of spin-1/2 fermions with resonant
interactions, as the one in Ref.~\cite{Sheehy2005}, is still open.
The main difficulty seems to be the presence of the spin, which
could complicate the Bethe Ansatz solution or even make it
impossible to construct. We leave this question for future work.

In this paper we construct solution to the problem defined in
\rfs{eq:FA} using Bethe Ansatz approach originated in
Ref.~\cite{Lieb1963}. These authors study \rfs{eq:FA} with $g=0$, so
we extend their techniques to the $g \not =0$ case. It is important
to stay away from real bound states of atoms in our construction, to
avoid the collapse of the gas. The condition for the absence of real
bound states in \rfs{eq:FA} (we will show it later in the paper)
reads
\begin{equation} \label{eq:nbs} \epsilon_0-\frac{g^2}{c}>0. \end{equation}
Thus we study \rfs{eq:FA} supplemented by this condition.

The bosonic Pauli principle for one dimensional interacting bosons
of the sort studied in Ref.~\cite{Lieb1963} (that is, \rfs{eq:FA}
with $g=0$) demands that no two bosons can asymptotically occupy the
same state. As a result, the bosonic system \rfs{eq:FA} at $g=0$ can
be described by the density of states $\rho(\lambda)$, where
$\lambda$ is the asymptotic momentum (momentum when the particles
are far away from each other). The ground state corresponds to all
the states up to the Fermi momentum $\pm \lambda_F$ filled, and the
rest empty. The value of $\lambda_F$ depends on the total density
$D$ of system (number of particles per unit length), with
$D(\lambda_F)$ being a monotonously growing function of its
argument. The excitation spectrum $\epsilon(\lambda)$ can also be
constructed \cite{Lieb1963a}. It is a monotonously growing function
of $\lambda$ at $\lambda>0$, negative at $|\lambda|<\lambda_F$ and
positive at $|\lambda|>\lambda_F$. It describes particle and hole
excitations. In the vicinity of $\lambda_F$ it can be replaced by a
linear function of $\lambda$ well described by the Luttinger liquid
theory.

We show that the Pauli principle also holds for the Feshbach
Hamiltonian \rfs{eq:FA} when applied to the atoms \cite{zeroener}.
However, the excitation spectrum $\epsilon(\lambda)$ is more
complicated than at $g=0$. Three regimes of the resonant Hamiltonian
can be identified, depending on the range of values of $c$. These
are
\begin{equation} \label{eq:regimes}
1.~ c \ll \sqrt{\frac{\epsilon_0}{m_a}}, \ 2. ~c \sim
\sqrt{\frac{\epsilon_0}{m_a}},  \ 3. ~c \gg
\sqrt{\frac{\epsilon_0}{m_a}}.
\end{equation}
In all three regimes, $g$ is assumed to be in the range such that
\rfs{eq:nbs} holds.

If the total density $D$ is small, so that $\lambda_F$ is much
smaller than the resonant momentum (momentum where an atom
scattering off another atom with opposite and equal  momentum would
be in resonance with the diatomic molecule), then the density of
states $\rho(\lambda)$ and the excitation spectrum
$\epsilon(\lambda)$ practically coincides with that at $g=0$ (delta
function interacting Bose gas) studied in Ref.~\cite{Lieb1963} and
\cite{Lieb1963a}. However, as the density of the gas is increased,
the density of states and the excitation spectrum develop new
features.

In the simplest of the three regimes, where $c \sim
\sqrt{\epsilon_0/m_a}$, the excitation spectrum remains roughly
similar to that of the delta function interacting Bose gas, even at
higher density. However, as $D$ is increased further, $\lambda_F$
can never exceed a certain cutoff momentum $\lambda_F^{\rm cutoff}$.
As $\lambda_F$ approaches the cutoff momentum, the gas accommodates
the higher density by increasing its density of states
$\rho(\lambda)$, as opposed to by increasing its Fermi momentum
$\lambda_F$. As a function of $\lambda$, the density of states
remains roughly constant in this regime. $\lambda_F^{\rm cutoff}$
can be calculated in a closed form, but the expression for it is not
simple. It can be estimated to be $ \lambda_F^{\rm cutoff} \sim  m_a
c$.

In the regimes of small or large $c$, the behavior deviates even
more from that of the delta function interacting gas. As the density
is increased,  $\epsilon(\lambda)$ ceases to be a monotonous
function of $\lambda$. Instead, it develops a local minimum above
the Fermi point at a momentum such that the particles moving at this
momentum are roughly at resonance with the particles in the Fermi
sea. (see Fig.~\ref{Fig:eps2a}, which depicts the case where $c$ is
small. The function plotted there is $\t \epsilon(\t
\lambda)=\epsilon/\lambda_F^2$ where $\t
\lambda=\lambda/\lambda_F$.) This minimum cannot be captured by
Luttinger liquid theory as it develops a finite distance above the
Fermi momentum and not in its immediate vicinity.

As the density is increased further, the minimum in
$\epsilon(\lambda)$ gets lower.  At a certain critical density
$D_{\rm crit}$ the minimum touches zero, and above this density the
minimum is now negative, Fig.~\ref{Fig:eps3a}. At this point the
ground state which resulted from filling the states with $\lambda$
between $-\lambda_F$ and $\lambda_F$ becomes unstable and the states
around the minimum also start getting filled. The actual value of
the critical density can be found analytically only if $c$ is small
or $c \ll \sqrt{\epsilon_0/m_a}$, and it is given by
\begin{equation} \label{eq:Dcrit0}
D_{\rm crit} = \left( \frac{ 9 m_a \epsilon_0^3}{2\left(c \epsilon_0
- g^2 \right)} \right)^{\frac{1}{3}}.
\end{equation}
The denominator of \rfs{eq:Dcrit0} is positive due to \rfs{eq:nbs}.
The Fermi momentum of the critical gas is given by, again at small
$c$
\begin{equation} \label{eq:lcrit0}
\lambda_F =  \left( 6 m_a^2 \left(c \epsilon_0 - g^2 \right)
\right)^\ot.
\end{equation}
The case of large $c$ is harder to study analytically, so we do not
have analytic equivalents of Eqs.~\rf{eq:Dcrit0} and \rf{eq:lcrit0},
but on the qualitative level the behavior of the excitation spectrum
is similar in both cases.

Unfortunately, we are unable to determine precise properties of the
ground state and excitations once the system crosses over into this
new high density regime. This is due to the fact that the equation
which describes the new ground state is nonlinear, and it is not
clear how to go about solving it, either analytically or
numerically. So in this paper we limit ourselves with speculations
about the properties of this high density state.


Which of the three regimes can be realized in the experiment depends
on the properties of a particular Feshbach resonance. This is an
important question, but we will not attempt to study it in this
paper.

Finally, we address briefly a possible criticism which can be
brought against the technique we use in this paper. This technique,
taking into account merely pair-wise interactions between the atoms,
does not explicitly take into account that the atoms spend some part
of their time bound into molecules, and that a molecule can interact
with another atom in its vicinity. We observe, however, that the
density of the molecules is very small, at least in the small and
large $c$ regimes and at total densities $D$ not exceeding too much
the critical density at which the phenomena discussed in this paper
take place. This is due to the fact that few atoms are in resonance
with each other in the regimes of interest here. Thus interactions
between the molecules, in the regimes of interest here, can be
neglected. We discuss this in more detail at the end of
Section~\ref{sec:BA} and in the Appendix~\ref{AppendixF}.


The rest of the paper is organized as follows. The Bethe Ansatz
techniques require the knowledge of the two-body scattering matrix
$S$. Section \ref{scattheory} is devoted to constructing the
scattering theory in one dimension and applying it to \rfs{eq:FA} on
the two-body level. This section can be skipped at a first reading,
and its only result important for further applications is the
two-body scattering matrix \rfs{eq:basicS} for the Hamiltonian
\rfs{eq:FA}.

In the next section \ref{sec:BA} we construct Bethe Ansatz many body
wavefunctions out of the two-body scattering matrix $S$. The
procedure is fairly standard and is included for completeness.

The next section \ref{sec:BE} is devoted to the Bethe equations
which allow to compute the density of states and the excitation
spectrum of the Bose gas. Although standard, these equations involve
the scattering phaseshift $\delta$, related to the $S$-matrix as in
$S=e^{2 i \delta}$. The phase shift for resonantly interacting gas
takes a new form which follow from \rfs{eq:basicS}. Thus in the
remainder of that section we discuss qualitatively what behavior we
can expect from $\delta$, identifying the physically distinct large
$c$ and small $c$ limits, depicted on Figs.~\ref{Fig:theta2} and
\ref{Fig:theta3}. The functions $\theta(\lambda)$ shown on these
figures is related to $\delta$ by \rfs{eq:deftheta}.

Armed with these preliminaries we proceed to construct the ground
state and the excitations of \rfs{eq:FA} in section \ref{sec:GSE}.
It is in this section that the main results of this paper are
derived and discussed.

Finally, in  section \ref{sec:BSGS} we discuss possible structure of
the ground state and excitations at densities above the critical
density.

Finally, several appendices discuss technical details relevant for
the material presented in this paper.

One of the most detailed discussions in the literature of the
standard one dimensional interacting Bose gas, without resonances,
is presented in the book Ref.~\cite{KorepinBook}. Throughout this
paper we extensively borrow their notations and use their results
when needed.

\section{One Dimensional Scattering}
\label{scattheory}

We begin this paper with constructing the two-body $S$-matrix for
the Hamiltonian \rfs{eq:FA}, which will be of use for the subsequent
solution of the problem using Bethe Ansatz. Given a lack of
discussions in the literature, it seems appropriate not only to
construct it for this specific Hamiltonian, but also to discuss its
general features, which is the subject of this section of the paper.

Quantum mechanics textbooks typically spend substantial time
discussing scattering in three dimensional space. At the same time,
scattering in one dimension is discussed only briefly. One
dimensional scattering is in many regards similar to the one in
three dimensions, yet there are also crucial differences.

Consider a particle in one dimension moving in a symmetric
potential $U(x)=U(-x)$ which quickly vanishes at large distances,
$U(x) \rightarrow 0$ if $|x| \gg r_0$,
\begin{equation} \label{eq:1dsh}
-\frac{1}{2m} {d^2 \psi \over dx^2} + U(x) \psi = E \psi.
\end{equation}
Here, as throughout this paper, the Planck constant $\hbar$ is set
to 1.

Typically for a problem defined in \rfs{eq:1dsh} we set up a
transmission-reflection problem. It is defined as a solution to
\rfs{eq:1dsh} with the asymptotic behavior of $\psi(x)$ given by
\begin{eqnarray}
\psi(x) = e^{i k x} + r e^{-i k x},&&  x \ll -r_0, \cr \psi(x) = t
e^{i k x}, && x \gg r_0.
\end{eqnarray} Here $r$ and $t$ are transmission and reflection
amplitudes, with transmission and reflection probabilities given
by $T=|t|^2$, $R=|r|^2$. However, for a symmetric potential, it is
advantageous to set up a scattering problem defined in a way
analogous to the scattering theory in 3D,
\begin{eqnarray} \label{eq:scatamp}
\psi(x)=e^{i k x} + f_S e^{-i k x}-f_A e^{-i k x}, && x \ll -r_0,
\cr \psi(x) = e^{i k x} + f_S e^{i k x} + f_A e^{i k x}, && x \gg
r_0.
\end{eqnarray}
Here $f_S$ and $f_A$ are symmetric and antisymmetric scattering
amplitudes (similar to the partial wave scattering amplitudes in
three dimensions). It is clear that $r=f_S-f_A$, and
$t=1+f_S+f_A$. Another way to set up a scattering problem is via
an S-matrix. It is defined via a solution to \rfs{eq:1dsh} with
the asymptotic behavior
\begin{eqnarray} \label{eq:sw}
\psi(x)=e^{i k x} + S_S e^{-i k x}, && x \ll -r_0, \cr \psi(x) =
e^{-i k x} + S_S e^{i k x}, && x \gg r_0
\end{eqnarray}
for the scattering in the symmetric channel, and
\begin{eqnarray} \label{eq:asw}
\psi(x)=-e^{i k x} + S_A e^{-i k x}, && x \ll -r_0, \cr \psi(x) =
e^{-i k x} - S_A e^{i k x}, && x \gg r_0
\end{eqnarray}
in the antisymmetric channel. By constructing $\psi(x)+\psi(-x)$
and $\psi(x)-\psi(-x)$ out of \rfs{eq:scatamp} it is easy to see
that
\begin{equation}
f_S=\frac{S_S - 1}{2}, \ f_A=\frac{S_A-1}{2}.
\end{equation}
By probability conservation the scattering matrix is unitary. As a
result, the scattering matrix can be represented in terms of phase
shifts
\begin{equation}
S_S = e^{2 i \delta_S}, \ S_A = e^{2 i \delta_A}.
\end{equation}
Another convenient representation of the scattering amplitude
takes the form
\begin{equation} \label{eq:invform}
f_S=\frac{1}{ik F_S(k^2)-1}, \ f_A=\frac{1}{i \frac{
F_A(k^2)}{k}-1}.
\end{equation} Here
$F_S(k^2)$ and $F_A(k^2)$ are real functions of its argument,
Taylor expandable in powers of $k^2$ at small $k$. The fact that
${\rm Re}(f^{-1}_{S,A})=-1$ is simple to establish. It follows
immediately from the unitarity of the S-matrix, or from $|1+2
f_{S,A}|^2=1$. It is somewhat more difficult to establish the form
of ${\rm Im}(f^{-1}_{S,A})$, this is done in Appendix
\ref{AppendixA}. All of these results closely parallel well known
facts from the theory of 3 dimensional scattering (see, for
example, \cite{LL}).


In this paper,
we are interested in scattering of bosons, which, due to
the symmetry under particle exchange, do not scatter at all in the
antisymmetric channel.  So, we will not study scattering in the
antisymmetric channel any further. To simplify notations, in what
follows we will drop the subscripts $S$ and $A$ and use the
notation such as $f\equiv f_S$, $S\equiv S_S$, and $F\equiv F_S$.

If we are interested in scattering at sufficiently low energy we
can replace the function $F(k^2)$ by its value at zero momentum,
$F(0)=-a$. Here $a$ is usually called the scattering length. Thus
the low energy scattering amplitude and S-matrix take the form
\begin{equation} \label{eq:zes}
f=-\frac{1}{i a k+1}, \ S=\frac{i a k -1}{i a k+1}.
\end{equation}
It is important to keep in mind that it is not enough to take $k
\ll r_0^{-1}$ for this simplification to work. In fact, the
condition specifying how small $k$ must be for \rfs{eq:zes} to
work depends on the details of the potential. In all the examples
of the potentials of interest in this paper, $k$ will be taken
much smaller than $r_0$, yet one will often have to go far below
that scale for \rfs{eq:zes} to hold. Yet, for arbitrary potential
there is always some scale so that below it \rfs{eq:zes} is
applicable \cite{zeroener}.

Notice that in the limit $k \ll a^{-1}$, the scattering amplitude
becomes $f=-1$, and the S-matrix is $S=-1$. This means that the
transmission coefficient, $t=1+f$, vanishes at low energy, while
the reflection coefficient $r=f=-1$. Thus we arrived at a result
of utmost importance in this paper: at very low energies the
scattering in an arbitrary potential in one dimension results in
total reflection (with one exception, see \cite{zeroener}). Stated
in this form, this is well known in quantum mechanics literature,
see, for example, Ref.~\cite{LL}, problem 5 in section 25.

The scattering phase shift is related to the function $F$ via a
simple formula
\begin{equation} \label{eq:lesp}
\cot \delta = - k F(k^2) \approx k a,
\end{equation} where the last approximate equality holds at sufficiently low momenta.
Thus at $|k a| \ll 1$, $\delta=\frac{\pi}{2}$ if $a$ is positive and
$\delta=-\frac{\pi}{2}$ if $a$ is negative (we restrict $\delta$ to
the interval $-\frac{\pi}{2}$ to $\frac{\pi}{2}$). Either way, this
means that scattering at very low energies is always in the {
unitary limit}. This result is specific to one dimension and has no
equivalent in three dimensional scattering.

An important potential to study is the delta-function potential
$U(x)=c~\delta(x)$, where $c$ controls its strength. It is
attractive at $c<0$ and repulsive at $c>0$. Let us calculate
scattering amplitude in this potential. It is easiest to do by
first computing the phase shift. We look for a solution of the
Schr\"odinger equation
\begin{equation}
-\frac{1}{2m} {d^2 \psi(x) \over dx^2} + c~ \delta(x)
\psi(x)=\frac{k^2 }{ 2} \psi(x)
\end{equation}
in the form $\psi(x)=\cos(kx+\delta)$ for $x>0$ and
$\psi(x)=\cos(-kx+\delta)$ for $x<0$ (this coincides with
\rfs{eq:sw} up to an overall phase, and of course, in the case of
delta-function, $r_0=0$). Direct substitution gives
\begin{equation}\label{eq:dfps}
\delta = - \arctan \frac{m c}{k}.
\end{equation}
The scattering amplitude and S-matrix become
\begin{equation} \label{eq:dfsa}
f=\frac{1}{\frac{i k }{m c}-1}, \ S=\frac{i k + m c}{i k -m c}.
\end{equation}
Notice that at $k=-i m c $ the scattering matrix has a pole. This
pole corresponds to the bound state with the energy $E=k^2/2m=- m
c^2 /2 $ if $c<0$ and the potential is attractive. If, however,
$c>0$, then this pole does not correspond to the bound state and
instead describes what is called in the literature a virtual bound
state \cite{LL}. The existence of the bound state at $c<0$, and
its absence at $c>0$, can also be verified by the direct
substitution of $k=-im c$ into \rfs{eq:sw}.

The result expressed by \rfs{eq:dfsa} is quite remarkable. It
implies that the scattering amplitude in the delta-function
potential coincides with the low energy asymptotics of the
scattering amplitude in arbitrary potential, with the
identification $m c=-1/a$. This is why when studying  many body
interacting systems with typical energies low enough for the
approximation leading to \rfs{eq:zes} to hold, it is sufficient to
replace the realistic interactions by the delta-function
interactions, which are often easier to analyze. And this is why
exact results obtained in Ref.~\cite{Lieb1963,Lieb1963a} for the
delta-function interacting one dimensional Bose gas are applicable
for the whole range of bosonic systems in one dimension regardless
of their interactions.

\begin{figure}[b]
\includegraphics[height=1.5in]{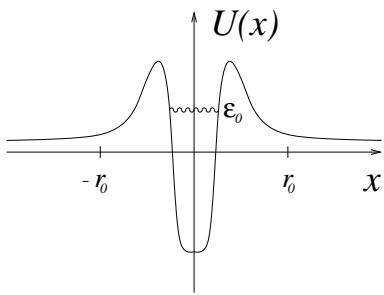}
\caption{\label{Fig:Pot} A symmetric potential $U(x)$ of the
typical range $r_0$ with the quasistationary energy level at some
energy $\epsilon_0$.}
\end{figure}

In this paper, however, we will be interested in potentials which
support a long lived quasistationary state (resonance) at positive
energy $E=\epsilon_0$. A potential of this kind is depicted on
Fig.~\ref{Fig:Pot}.

At energies far below $\epsilon_0$, the scattering amplitude of
this potential reduces to \rfs{eq:zes}, just like for any other
potential. However, at energies of the order $\epsilon_0$ the
scattering amplitude no longer takes the simple form \rfs{eq:zes}.
Indeed, the scattering amplitude must reflect in its pole
structure the presence of the resonance at $\epsilon_0$. That
means, it must have a pole at the complex value of the energy
\begin{equation} \label{eq:width} \frac{k^2}{2 m}=\epsilon_0-i \Gamma/2,
\end{equation} where $\Gamma$ is the inverse lifetime of the resonance
\cite{LL}.

Notice that the potential depicted on Fig.~\ref{Fig:Pot} can be made
arbitrarily narrow (and correspondingly deep, to keep $\epsilon_0$
at a finite value), so $\epsilon_0$ can always be made much smaller
than $1/m r_0^2$. Thus momenta of the order of $k \sim \sqrt{m
\epsilon_0}$ can be made much smaller than $r_0^{-1}$. Yet
\rfs{eq:zes} may already not apply. Instead we have to go back to a
more general result \rfs{eq:invform}.

To work out a typical scattering amplitude for a potential such as
the one depicted on Fig.~\ref{Fig:Pot} is not easy. Fortunately, we
do not need to work with an explicit potential. Instead, we recall
that we are interested in atoms interacting via a Feshbach
resonance, described by \rfs{eq:FA}. Two-body version of the
Feshbach Hamiltonian, sometimes referred to as a Fano-Anderson model
\cite{Fano1961} or a single level in the continuum \cite{MahanBook},
is
\begin{eqnarray} \label{eq:Fano}
H &=& \frac{1}{2m} \int dx~ \d_x a^\dagger \, \d_x a + \cr &+ & g
\left[ a^\dagger(0) b + b^\dagger a(0) \right]+\epsilon_0
b^\dagger b + c ~a^\dagger(0) a(0).
\end{eqnarray}
Notice that the numerical coefficients, present in \rfs{eq:FA} due
to indistinguishability of the particles, disappeared.  The
Fano-Anderson Hamiltonian describes a particle, created by
$a^\dagger$, moving a long a one dimensional line which, upon
hitting the point $x=0$, may turn into another particle, created by
$b^\dagger$. The amplitude of this process is given by $g$. The
$b$-particle does not move and has a fixed energy $\epsilon_0$. This
particle is unstable and after some time will turn back into an
$a$-particle which will then be free to move away from the origin.
In addition, we add a delta-function repulsion from the origin for
the $a$-particle, with strength $c$. As discussed in the
introduction, in one dimension the presence of the delta-function
repulsion makes a crucial difference (eliminates ``parasitic" bound
states), as we will see below.

It is not unreasonable to expect that this model describes an
effective potential with long lived resonance whose energy is
approximately $\epsilon_0$ (assuming that $\epsilon_0>0$). The
lifetime of the resonance is controlled by g. If $\epsilon_0<0$,
then we expect that it describes a potential with a bound state. We
will see below that this picture is more or less correct (although
the energy of the resonance is $\epsilon_0$ at small $c$ only. At
large $c$, it is at $\epsilon_1$, defined in \rfs{eq:epsilon1}).

To understand the Fano-Anderson model deeper, let us compute the
scattering amplitude of the $a$-particles. The scattering amplitude
is given by the $T$-matrix, which is closely related to the
scattering amplitude. In one dimension, this relation is given by
\begin{equation} \label{eq:ts}
f(k)=-{i m\over k} T(k).
\end{equation}
$T$ matrix itself, for the Hamiltonian \rfs{eq:Fano}, can be
computed diagrammatically. Leaving the details of the calculation
to the Appendix \ref{AppendixB}, we present the answer
\begin{equation} \label{eq:sc}
f(k)=\left( \frac{ik}{m} \cdot \frac{ \frac{k^2}{2m}-\epsilon_0}{
c \left( \frac{k^2}{2m}-\epsilon_0 \right) + g^2}-1\right)^{-1}.
\end{equation}

First of all, we see that this scattering amplitude conforms to a
general form \rfs{eq:invform} a symmetric scattering amplitude in
one dimension must assume.

Second, we observe that in the absence of Feshbach term, $g=0$,
this scattering amplitude reduces to scattering in the
delta-function potential, \rfs{eq:dfsa}, as it should.

Third, we would like to study this scattering amplitude  in the
absence of delta-function term, $c=0$. The scattering amplitude
reduces to a simpler form
\begin{equation} \label{eq:scv}
f_{c=0}(k) = \left( \frac{i k}{m g^2}  \left( \frac{k^2}{2m
}-\epsilon_0 \right) -1 \right)^{-1}.
\end{equation}
This indeed describes scattering via a potential with a resonance
close to $\epsilon_0$ (assuming it is positive) with the lifetime
controlled by $g$. By analyzing the poles of \rfs{eq:scv} it is
easy to see that one of them, if $g^2 \ll
\epsilon_0^{\frac{3}{2}}/\sqrt{m}$, is given by
\begin{equation}
\label{eq:res} \frac{k^2}{2m} \approx \epsilon_0-i  g^2
\sqrt{\frac{ m} {2 \epsilon_0}}.
\end{equation}

 From here and from \rfs{eq:width} we
deduce that the lifetime of the resonance is inversely
proportional to $g^2$.

However, the scattering amplitude \rfs{eq:scv} also harbors in
itself a bound state at arbitrary values of $\epsilon_0$. Indeed,
if $\epsilon_0$ is sufficiently large, then the energy of the
bound state is simply given by $k^2/2m \approx - m g^4/2
\epsilon_0^2$, where we balanced the term linear in $k$ in
\rfs{eq:scv} with $-1$, neglecting the cubic in $k$ term. If
$\epsilon_0$ is lowered until it becomes large negative, then this
bound state smoothly crosses over into a more obvious bound state
at $k^2/2m=\epsilon_0$, while the resonance which existed at
positive values of $\epsilon_0$ simply disappears (mathematically,
its pole moves off into an unphysical part of the complex energy
plane). This can be seen by finding the corresponding poles
exactly.

The existence of the bound state of \rfs{eq:Fano} at $c=0$ even when
$\epsilon_0$ is positive and large can also be understood by noting
that the $a$-particles, if their energy is below $\epsilon_0$, can
only spend a very short time in the $b$-state. Their scattering will
then be dominated by the second order perturbation theory where a
particle jumps into a highly energetic $b$-state with amplitude $g$,
and then jumps back down to one of the $a$-states. Such process
always leads to effective attraction between $a$-states and the
origin. Unlike the three dimensional space, in one dimension an
attraction, no matter how weak, leads to a bound state. That's the
bound state we observe here.

A system of one dimensional bosons with pair-wise short range
interactions which support bound states is unstable and its energy
per particle in the thermodynamic limit actually becomes
infinitely negative. (Ref.~\cite{Lieb1963}. See also discussion in
the next section.) That's why we would like to study potentials
when there are no bound states.

Fortunately, the full problem \rfs{eq:Fano} with additional
repulsive interaction $c>0$, whose scattering amplitude is given by
\rfs{eq:sc}, does not have any bound states as long as the condition
\rfs{eq:nbs} is fulfilled. Yet it does possess a resonance.
Introducing notation
\begin{equation} \label{eq:epsilon1} \epsilon_1=\epsilon_0-\frac{g^2}{c},\end{equation}
we rewrite the
scattering amplitude as in
\begin{equation} \label{eq:basic}
f(k)=\left( \frac{ik}{m c} \frac{ k^2-2 m \epsilon_0}{ k^2-2 m
\epsilon_1}-1\right)^{-1}.
\end{equation}
Here $\epsilon_0>\epsilon_1>0$. Indeed, at small $c$ and $g$, this
amplitude has a pole at $ k^2/2 m\approx \epsilon_0-i g^2 \sqrt{m/2
\epsilon_0}$, same as \rfs{eq:res}. At large $c$ and $d$, the pole
is at $k^2/2m \approx \epsilon_1-i (g^2/c^2) \sqrt{2 \epsilon_1/m}$.
In either of the two cases, the lifetime of the resonance is large.

We take the scattering amplitude \rfs{eq:basic} as the basic
resonant scattering amplitude of interest in this paper. For
future reference, we also write down the S-matrix corresponding to
the amplitude \rfs{eq:basic}
\begin{equation} \label{eq:basicS}
S(k)=2 f(k)+1=\frac{ \frac{ik}{m c} \frac{ k^2-2 m \epsilon_0}{
k^2-2 m \epsilon_1}+1}{ \frac{ik}{m c} \frac{ k^2-2 m \epsilon_0}{
k^2-2 m \epsilon_1}-1}.
\end{equation}

The two limits of small or large $c$ and $g$, where the problem
possesses a narrow resonance, have an interesting interpretation. In
one of the regimes, where $c \ll \sqrt{\epsilon_0/m}$, we observe
that $f(k)$ is very small almost everywhere except in the vicinity
of $k=0$ and $k^2/2m=\epsilon_0$, where $f \approx -1$. Recalling
that the transmission amplitude $t=1+f$, we find that this regime
corresponds to perfect transmission at all values of momentum except
at very low momenta and at the resonance, where we have total
reflection. This is the regime 1 of \rfs{eq:regimes}.

The other regime, $c \gg \sqrt{\epsilon_0/m}$, the regime 3 of
\rfs{eq:regimes}, leads to $f(k)=-1$ almost everywhere except at
$k^2/2 m=\epsilon_1$, where $f=0$. Thus it describes a system with
total reflection at all values of energy except at resonance (which
is at $\epsilon_1$), where we have total transmission. It is not
unreasonable to expect that this situation describes the
transmission/reflection amplitude in case of the potential depicted
on Fig.~\ref{Fig:Pot} (with $\epsilon_1$ playing the role of
$\epsilon_0$ on the figure).

Whether any of  the two limits is appropriate to describe
experimentally observed Feshbach resonances depends on the
properties of a particular resonance. This question deserves further
study, but will not be discussed in this paper.

\section{Bethe Ansatz Solution to the Many-Body Schr\"odinger equation}
\label{sec:BA}

Consider a gas of bosons moving on a line and interacting via a
pair-wise interaction $U(x)$
\begin{equation} \label{eq:mbs}
- \oh  \sum_j^N \frac {\d^2 \psi} {d x_j^2} + \sum_{N \ge j>k\ge1}
U(x_j-x_k) = E \psi.
\end{equation} The wave function $\psi(x_1,x_2,\dots)$ must be
symmetric under interchange of any of the coordinates. For
simplicity, we set the mass of bosons to be equal to $m_{a}=1$;
hence the coefficient $1/2$ in front of the second derivative. Note
that the reduced mass of a pair of bosons is now set at $$m=1/2,$$
so all the scattering formulae of the previous chapter should be
understood as having this mass.

Generally speaking, for a large number of particles $N$ and
arbitrary potential $U(x)$ it is not possible to solve this
equation. However, if $U(x)$ is very short ranged, then the
probability that three particles simultaneously interact with one
another is very low, and we can assume that only two particles
interact with each other at a given time. In this case, we can use
Bethe Ansatz to construct many body wave function. In what
follows, we will closely follow Ref.~\cite{KorepinBook} in
applying Bethe Ansatz to this problem, emphasizing the differences
whenever they are present.

Usually one constructs these wave functions for the delta function
interactions \cite{KorepinBook}. However,  as we saw in the previous
section, there are other  potentials which are, on the one hand,
short ranged, and on the other hand, have interesting structure. The
potential depicted on Fig.~\ref{Fig:Pot} is one example if it is
made sufficiently short ranged ($r_0 \ll 1/k$ and also much less
than the typical spacing between the bosons). Even better example is
given by the particles interacting via Feshbach resonance
\rfs{eq:FA}. Just as in \rfs{eq:mbs}, in what follows we take the
mass of the bosons $m_a$ in  \rfs{eq:FA} to be $1$, which means that
the mass $m$ in the two-body scattering matrix $S$ is now equal to
the reduced mass of the bosons, or $1/2$.

 Bethe
Ansatz for the many body wave function which solves \rfs{eq:mbs}
or \rfs{eq:FA} can be written down in the following manner. First
we construct an unsymmetrized wave function $\t \psi$
\begin{widetext}
\begin{eqnarray}
\label{eq:perm} \t \psi(x_1,x_2,\dots) &=& e^{i \sum_j \lambda_j
x_j}, \ x_1<x_2<\dots<x_N, \cr \t \psi(x_1,x_2,\dots) &=& e^{i
\sum_j \lambda_j x_j} S\left(\frac{\lambda_1-\lambda_2}{2}
\right), \ x_2<x_1<x_3<\dots<x_N, \cr \t \psi(x_1,x_2,\dots) & = &
e^{i \sum_j \lambda_j x_j}
S\left(\frac{\lambda_1-\lambda_2}{2}\right)
S\left(\frac{\lambda_1-\lambda_3}{2} \right), \
x_2<x_3<x_1<x_4<\dots<x_N, \cr \dots &&
\end{eqnarray}
\end{widetext}
Here $S(\lambda)$ is the scattering matrix in the potential $U(x)$,
computed by solving the Schr\"odinger equation \rfs{eq:1dsh} with
$m$ being the reduced mass of the particles, $m=1/2$. To write
\rfs{eq:perm} down, we add a factor of $S((\lambda_j-\lambda_k)/2)$
for every elementary permutation between $x_j$ and $x_k$ ($j<k$) one
needs to do to rearrange the list $x_1,x_2,x_3,\dots$ in the order
of increasing $x$, $x_{j_1}<x_{j_2}<\dots$.

Then we symmetrize the function $\t \psi$ to arrive at the
symmetric wave function
\begin{equation} \label{eq:sol}
\psi=\sum_P \t \psi(x_{P_1},x_{P_2},\dots),
\end{equation}
where $P_j$ represents permutations of $1,2 \dots, N$. $\psi$
satisfies \rfs{eq:mbs} with the energy \begin{equation}
\label{eq:energy} E= \oh \sum_j \lambda_j^2. \end{equation} This
is proven in Appendix \ref{AppendixC}. This wave function is a
generalization of a standard Bethe Ansatz construction for the
many body problem \rfs{eq:mbs} with the delta-function potential
\cite{KorepinBook}.

One notable property of this wave function lies in the following:
if any two $\lambda$ happen to coincide, $\psi=0$. Indeed, suppose
$\lambda_j=\lambda_k$. Then it is straightforward to see that $\t
\psi(x_1,x_2,\dots)$ is antisymmetric under interchanging $x_j$
and $x_k$. Two crucial properties of the S-matrix necessary to
prove that are $S(0)=-1$ and $S(\lambda) S(-\lambda)=1$, and the
proof is discussed in more detail in Appendix \ref{AppendixC}.
Thus $\psi$, which is a symmetrized version of $\t \psi$,
vanishes. Therefore, all $\lambda_k$ must be different for the
solution to the Schr\"odinger equation to exist.

This property of the many body one-dimensional Schr\"odinger
equations for bosons is sometimes referred to as a bosonic ``Pauli
principle". Interacting bosons in one dimension must all have
different momenta, as if they are fermions. While discussed in
textbooks for delta-function potential \cite{KorepinBook}, this
property holds for arbitrary short ranged potentials (with one
exception \cite{zeroener}).

Finally let us discuss the importance of the 3-body processes
neglected here. We observe that in the process of checking that the
wave function \rfs{eq:sol} satisfies the Shr\"odinger equation, we
neglected the following 3-body processes. Suppose two of the
coordinates, say $x_j$ and $x_k$, are taken to be equal to each
other. And suppose a third coordinate $x_l$ is passed through this
point. Then \rfs{eq:sol} may violate the Schr\"odinger equation in
this range of coordinates.

When the interactions are given by delta function, then the
integrability protects the 3-body processes from violating the
Schr\"odinger equation. More generally, if the interactions are
short ranged, then the weight of the wave function when two points
are within the range of the interactions $r_0$ is negligibly small,
and 3-body processes can be neglected. However, in our problem, due
to the presence of the resonance, we can imagine that the wave
function has a spike every time two particles in resonance approach
each other. Thus a substantial part of the weight of the wave
function can be at points where $x_j=x_k$.

Although one can imagine this may invalidate our solution
\rfs{eq:sol} when the density $D$ of the gas is so high that there
is a finite density of the molecules in the system, we show in the
Appendix~\ref{AppendixF} that in the interesting regimes of small
$c$ or large $c$, at the density of the gas around the point of
instability (given by \rfs{eq:Dcrit0} at small $c$, for example),
the density of the molecules is still sufficiently small so that we
can neglect 3-body processes and rely on \rfs{eq:sol} for
constructing solutions to our problem.

As the density of the gas is increased, the molecular density
increases as well, and at some point the techniques discussed in
this paper breaks down. However, this happens at densities well
above the ones where the phenomena discussed here take place.

Ideally it would be interesting to find a model, similar in its
physical properties to \rfs{eq:FA}, which would be integrable so
that the integrability would protect the Bethe Ansatz at all
densities. Whether this is possible is not known.

\section{Bethe Equations}
\label{sec:BE}

Not all sets of $\lambda_j$ are appropriate for the wave function
\rfs{eq:sol}. The wave function must satisfy appropriate boundary
conditions. As always, the specific form of the boundary conditions
is irrelevant while it is easiest to study periodic boundary
conditions. Note that if $\t \psi$ satisfies periodic boundary
conditions, then $\psi$ also satisfies them. Imposing periodic
boundary conditions on $\t \psi$, we find
\begin{equation} \label{eq:betheeq1}
L \lambda_j + \frac{1}{i}\sum_{k \not = j} \log
S\left(\frac{\lambda_j-\lambda_k}{2} \right) = 2 \pi \t n_j,
\end{equation}
where $\t n_j$ are arbitrary integer numbers, and $L$ is the size
of the system. It is customary in the Bethe Ansatz literature to
define a modified scattering phase shift $\theta$, related to the
standard phase shift $2i\delta=\log S$ in the following manner
\begin{eqnarray} \label{eq:deftheta}
\theta(\lambda)&  = & 2 \delta(\lambda)+\pi, \ \lambda>0, \cr
\theta(\lambda) & = & -\theta(-\lambda) , \ \lambda<0.
\end{eqnarray}
The scattering length of the S-matrix of interest to us
\rfs{eq:basic}, under condition \rfs{eq:nbs}, is negative (this is
related to the fact that the potentials of interest to us have no
bound states). For negative scattering length, $\lim_{\lambda
\rightarrow 0} \delta(\lambda)=-\pi/2$ (see discussion after
\rfs{eq:lesp}. Thus, we find that $\theta(0)=0$ and $\theta$ is an
antisymmetric continuous function. In terms of $\theta$,
\rfs{eq:betheeq1} becomes
\begin{equation} \label{eq:betheeq}
L \lambda_j + \sum_{k=1}^N
\theta\left(\frac{\lambda_j-\lambda_k}{2} \right) = 2 \pi n_j,
\end{equation}
where $n_j$ are integers if $N$ is odd, and $n_j$ are half
integers if $N$ is even. These equations are called Bethe
equations in the literature. They provide restriction on possible
values $\lambda_j$ may take. We need to solve them to find the
energy spectrum of our problem.

Let us recall several useful properties of Bethe equations. It is
shown in Ref.~\cite{KorepinBook} that all the solutions
$\lambda_j$ of the equations \rfs{eq:betheeq} are real provided
that
\begin{eqnarray} \label{eq:usS}
\left| S(\lambda) \right| \le 1 \ {\rm when} \ {\rm Im} ~\lambda
\ge 0, \cr \left| S(\lambda) \right| \ge1 \ {\rm when} \ {\rm Im}
~\lambda \le 0
\end{eqnarray}
Indeed, \rfs{eq:basicS}, together with \rfs{eq:nbs}, satisfies these
conditions. This can be shown with some straightforward algebra.
Thus, in the absence of bound states (condition expressed by
\rfs{eq:nbs}), all solution to the Bethe equations are real.

Second important property stems from the fact that
\begin{equation}\label{eq:monoto}
\dbyd{\theta(\lambda)}{\lambda}>0.
\end{equation} This is also
true curtesy of \rfs{eq:nbs}, as can be checked directly by
examining the explicit form of the derivative of $\theta$, given by
\rfs{eq:Klarge}. It is shown in Ref.~\cite{KorepinBook} that as a
consequence of positivity of derivative of $\theta$, the solutions
to the Bethe equations $\lambda_j$ can be uniquely parametrized by
the set of integers (half-integers) $n_j$. Conversely, in the
presence of bound states (violation of condition \rfs{eq:nbs}), the
function $\theta$ is no longer monotonous, so there could be
multiple solutions to Bethe equations.

Finally, it is shown in Ref.~\cite{KorepinBook} that under the
same condition \rfs{eq:monoto} if $n_j>n_k$, then
$\lambda_j>\lambda_k$. If $n_j=n_k$, then $\lambda_j=\lambda_k$.
Recall that if $\lambda_j=\lambda_k$, then the wave function
$\psi$ vanishes. It follows that no two $n_j$ may be the same.

To summarize, a set of distinct integers (half-integers) $n_j$,
under conditions of interest to us,  uniquely parametrizes the
allowed momenta $\lambda_j$ of the particles, and larger $n_j$
correspond to larger $\lambda_j$.

In the remainder of this section let us analyze the behavior of
$\theta$ of use in subsequent discussions in this paper.

\begin{figure}[tb]
\includegraphics[height=1.5in]{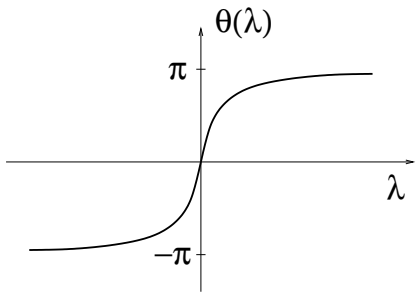}
\caption{\label{Fig:theta1} A plot of $\theta(\lambda)$ for a
delta function potential.}
\end{figure}

For a delta function potential, the phase shift is given by
\rfs{eq:dfps}. The phase $\theta$, constructed with the help of
\rfs{eq:deftheta} is depicted on Fig.~\ref{Fig:theta1}. It is a
monotonously increasing function, reaching $\pi$ at $+\infty$ and
$-\pi$ at $-\infty$.

\begin{figure}[tb]
\includegraphics[height=1.5in]{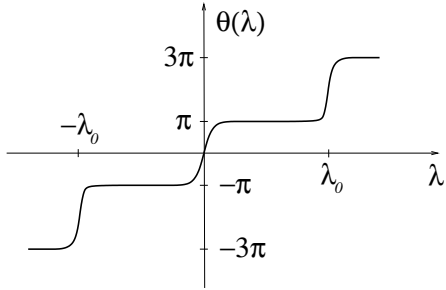}
\caption{\label{Fig:theta2} A plot of $\theta(\lambda)$ for a
resonant potential with scattering amplitude given by
\rfs{eq:basic}, and with $c$ small. Here $\lambda_0=\sqrt{2 m
\epsilon_0}$. }
\end{figure}

For a resonant scattering amplitude \rfs{eq:basic}, the form of the
phase $\theta(\lambda)$ depends on the two possible regimes,
discussed in text after \rfs{eq:basicS}. In the regime of small $c$,
it is schematically depicted on Fig.~\ref{Fig:theta2}. At
$|\lambda|\lesssim \lambda_0$, where a convenient notation is
introduced to be used extensively in what follows
\begin{equation} \lambda_0=\sqrt{2 m \epsilon_0}, \ \lambda_1=\sqrt{2 m \epsilon_1}=
\sqrt{2 m \left( \epsilon_0-\frac{g^2}{c} \right)},
\end{equation}
it closely follows the graph for $\theta(\lambda)$ for the delta
function potential. However, at larger $|\lambda|$ the phase jumps
by another $2 \pi$, characteristic of resonances in scattering
theory (the phase shift $\delta$ jumps by $\pi$). Notice that
$\theta(\pm \lambda_0)= \pm 2 \pi$. This  describes a situation
where at almost all the energies $\theta=\pi$ or $\theta=3\pi$,
which corresponds to the total transmission. In the vicinity of
$\lambda=0$ and $\lambda=\pm \lambda_0$, the phase goes though
$\theta=0$ or $\theta=\pm 2 \pi$, which is total reflection. The
total change in the phase $\theta(\infty)-\theta(-\infty)=6 \pi$.

\begin{figure}[tb]
\includegraphics[height=1.5in]{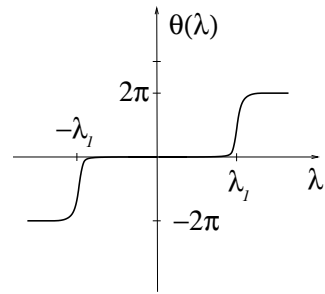}
\caption{\label{Fig:theta3} A plot of $\theta(\lambda)$ for a
resonant potential with scattering amplitude given by
\rfs{eq:basic} and with $c$ large. Here $\lambda_1=\sqrt{2 m
\epsilon_1}$.}
\end{figure}

If $c$ is large, then the phase $\theta$ is depicted on
Fig.~\ref{Fig:theta3}. The phase is either $0$ or $2 \pi$ almost
everywhere except in the vicinity of $\lambda=\pm \lambda_1$,
$\lambda_1=\sqrt{2 m \epsilon_1}$, where $\theta$ goes through
$\pi$. This corresponds to total reflection almost everywhere except
at energy equal to $\epsilon_1$. We note that at very large
$\lambda$, not depicted on Fig.~\ref{Fig:theta3}, $\theta$
subsequently increases to $3\pi$. However, it does it very slowly
and over a long interval of momenta (in the limit of very large
$c$). We will not consider behavior of $\theta$ at such large
momenta. So for our purposes $\theta(\infty)-\theta(-\8)=4 \pi$ in
this case.

Finally, we repeat that to conform with Eqs.~\rf{eq:mbs} and
\rf{eq:FA}, $m$ in the preceding paragraphs should be chosen as the
reduced mass of the particles there, or $m=1/2$.

In what follows, we will need the function
\begin{equation} \label{eq:kerneldef}
K(\lambda) = \dbyd{\theta\left(\frac{\lambda}{2}
\right)}{\lambda}.
\end{equation}

For delta function potential, this function can be easily computed,
by differentiating \rfs{eq:dfps}. The result is given by
\begin{equation} \label{eq:deltakernel}
K(\lambda) = \frac{2 c }{c^2+\lambda^2}.
\end{equation}

For the resonant amplitude \rfs{eq:basic} this function can also be
computed directly, with the result
\begin{equation} \label{eq:Klarge}
K(\lambda) = 2 c\frac{ \lambda^4+4 \left(\lambda_0^2-3
\lambda_1^2\right) \lambda^2+16
   \lambda_0^2 \lambda_1^2}{\left(\lambda^3-4 \lambda
   \lambda_0^2\right)^2+c^2 \left(\lambda^2-4
   \lambda_1^2\right)^2}.
\end{equation}
This formula is not very illuminating, except if used for numerical
calculations. Instead, we will concentrate on the cases of small or
large $c$. We notice that $K$, being the derivative of $\theta$
depicted on Fig.~\ref{Fig:theta2} and Fig.~\ref{Fig:theta3}, is zero
almost everywhere except in the vicinity of $2 \lambda_0$ and $2
\lambda_1$ respectively. Thus, it can be well approximated by
\begin{equation} \label{eq:kernel1}
K(\lambda) =\frac{2 c_1 }{c_1^2+\lambda^2}+\frac{2 c_2
}{c_2^2+\left(\lambda-2\lambda_0 \right)^2}+\frac{2 c_2
}{c_2^2+\left(\lambda+2\lambda_0 \right)^2}
\end{equation}
for small $c$, with
$$c_1=c \frac{\lambda_1^2}{\lambda_0^2}, \ c_2=c \frac{\lambda_0^2-\lambda_1^2}{2\lambda_0^2}.$$
$\theta$, which corresponds to \rfs{eq:kernel1}, is depicted on
Fig.~\ref{Fig:theta2}. For large $c$, the approximation reads
\begin{equation} \label{eq:kernel2}
K(\lambda) =\frac{2 c_3 }{c_3^2+\left(\lambda-2\lambda_1
\right)^2}+\frac{2 c_3 }{c_3^2+\left(\lambda+2\lambda_1 \right)^2},
\end{equation}
where $$c_3= 2 \frac{\lambda_0^2-\lambda_1^2}{c} ,$$ and the
corresponding $\theta$ is depicted on Fig.~\ref{Fig:theta3}.

If $c$ is neither small nor large, a surprisingly good approximation
amounts to replacing $K(\lambda)$ by its value of $\lambda=0$, or
\begin{equation} \label{eq:lacrude}
\theta(\lambda) \approx \frac{4 \lambda_0^2}{c \lambda_1^2} \lambda.
\end{equation}
This approximation, although crude and not extendable to large
$\lambda$ (since $\theta(\lambda) < 3 \pi$), nevertheless allows to
capture most features of the moderate $c$ regime rather well, as we
will see below.

\section{Ground State and Excitations of the Resonant Bose Gas}
\label{sec:GSE}

To construct the ground state of the system of $N$ particles
governed by the pair-wise S-matrix $S$, we need to choose a set of
$\lambda_j$, which satisfies Bethe equations \rfs{eq:betheeq}
while minimizing the energy \rfs{eq:energy}, with all $\lambda_j$
being different. It seems reasonable that for that we need  to
choose $n_j$ as small as possible. Together with the requirement
that no $n_j$ are the same, the ansatz for the ground state reads
\begin{equation} \label{eq:grn} n_j=-\frac{N-1}{2}, -\frac{N-1}{2}+1, \dots,
\frac{N-1}{2}.\end{equation} This ansatz indeed works for the delta
function potential \cite{KorepinBook}. {\sl It will be shown below
that in case of resonant scattering at small or large $c$,  this
ansatz breaks down as soon as the density of the gas exceeds a
certain critical value.} Nevertheless, we would like to work out the
consequences of \rfs{eq:grn} here, in the process obtaining the
criteria when \rfs{eq:grn} holds.

Following Ref.~\cite{KorepinBook}, we introduce function
$\lambda(x)$ such that $\lambda(n_j)=\lambda_j$, which satisfies,
buy virtue of \rfs{eq:betheeq}, the equation
\begin{equation}
L \lambda(x) + \sum_{k=1}^N \theta \left( \frac{\lambda(x) -
\lambda_k }{2} \right) = 2 \pi x.
\end{equation}
Differentiating this equation with respect to $\lambda$, and
introducing a function
\begin{equation}
\rho(\lambda) =\frac{1}{L} \dbyd{x}{\lambda},
\end{equation}
where $L$ is the length of the system, we find in the
thermodynamic limit $N \rightarrow \infty$
\begin{equation} \label{eq:lin}
\rho(\lambda)-\frac{1}{2 \pi} \int_{-\lambda_F}^{\lambda_F} d
\mu~K(\lambda-\mu) \rho(\mu) = \frac{1}{2 \pi},
\end{equation}
where $K$ was defined in \rfs{eq:kerneldef}.  \rfs{eq:lin} is called
the Lieb-Liniger equation \cite{Lieb1963,KorepinBook}, and $K$ can
be referred to as the kernel of this integral equation. Here
$\lambda_F=\lambda_N=-\lambda_1$, is the highest momentum the
particles in the ground state have. In this way, it is akin to Fermi
momentum, thus the notation. $\rho(\mu)$ has the meaning of density
of $\lambda_j$, with $\rho(\lambda_j)=1/L (\lambda_{j+1}-\lambda_j)$
in the thermodynamic limit. In other words, it is the density of
states per unit momentum. The total gas density can be expressed in
terms of $\rho$ as follows
\begin{equation}
D=\frac{N}{L}=\int_{-\lambda_F}^{\lambda_F} d \mu~\rho(\mu).
\end{equation}

By construction, $\rho(\mu)$ must be a positive function. It is
shown in Ref.~\cite{KorepinBook} that this implies that
$$\delta(\lambda-\mu)-\frac{1}{2 \pi} K(\lambda-\mu)$$ must be a positive
definite operator, with its smallest eigenvalue no smaller than
$1/(2 \pi\rho_{\rm max})$, where $\rho_{\rm max}$ is the maximum
value of $\rho(\lambda)$ on the interval between $-\lambda_F$ and
$\lambda_F$.

A necessary condition for this operator to be positive definite is
\begin{equation} \label{eq:conditionnec}
\int_{-\lambda_F}^{\lambda_F} d
\lambda~\int_{-\lambda_F}^{\lambda_F} d \mu \left[
\delta(\lambda-\mu)-\frac{1}{2\pi} K(\lambda-\mu) \right]>0
\end{equation}
This is equivalent to \begin{equation} \label{eq:nescond}
\frac{1}{\lambda_F } \int_0^{\lambda_F} d\lambda~\frac{
\theta(\lambda)}{\pi} < 1.
\end{equation}
For the delta function potential whose phase $\theta$ is depicted
on Fig.~\ref{Fig:theta1}, \rfs{eq:nescond} is always fulfilled no
matter what $\lambda_F$. Thus, $\lambda_F$ can be arbitrary. And
indeed, it is possible to show that $\lambda_F$ depends on $D$,
and the bigger the total density $D$ is, the bigger $\lambda_F$
is.

However, the situation is drastically different for the resonant
$\theta$, depicted on Figs.~\ref{Fig:theta2} and \ref{Fig:theta3}.
The scattering phase $\theta$ from Fig.~\ref{Fig:theta2} results
in the relation \begin{equation} \label{eq:limit1} \lambda_F
\lesssim \lambda_0.\end{equation} At the same time, the scattering
phase from Fig.~\ref{Fig:theta3} results in
\begin{equation} \label{eq:limit2} \lambda_F \lesssim 2 \lambda_1.\end{equation}
In case of moderate $c$, we can use \rfs{eq:lacrude} to find
\begin{equation}
\label{eq:limit3} \lambda_F \lesssim \frac{\pi c \lambda_1^2}{2
\lambda_0^2},
\end{equation}
from which the estimate quoted in the Introduction follows. As a
result, the Fermi momentum of the resonantly interacting Bose gas in
one dimension cannot be arbitrarily large. Increasing the density of
the gas $D$ increases the Fermi momentum $\lambda_F$. However, as
$\lambda_F$ approaches its limit \rfs{eq:limit1}, \rfs{eq:limit2},
or \rfs{eq:limit3} the density of states $\rho(\lambda)$ starts
increasing very fast, and $D$ diverges to infinity as $\lambda$
approaches its upper limit. This is confirmed by numerical solution
of \rfs{eq:lin}. We will see below, however, that only in case of
moderate $c$ this limit can be reached. In case of small or large
$c$, an instability develops in the ground state before the limit is
reached.

The excitations above
the ground state governed by the Lieb-Liniger equation \rfs{eq:lin}
are controlled by the function $\epsilon(\lambda)$, which obeys
\begin{equation} \label{eq:exc}
\epsilon(\lambda)-\frac{1}{2\pi} \int_{-\lambda_F}^{\lambda_F}
d\mu~K(\lambda-\mu) \epsilon(\mu) = \frac{\lambda^2}{2}-h.
\end{equation}
Here $h$ is the chemical potential  and $\lambda^2/2-h$ is the
energy spectrum in the absence of interactions. This equation must
be solved with the condition $\epsilon\left(\pm \lambda_F
\right)=0$. This condition fixes $h$ if density $D$ is given.

Once the function $\epsilon(\lambda)$ is known, the energy of an
excitation is given by $\epsilon$, while the momentum of this
excitation is given by $k$, where
\begin{equation} \label{eq:physmom}
k = \lambda+\int_{-\lambda_F}^{\lambda_F} d\mu~\theta
\left(\frac{\lambda-\mu}{2} \right) \rho(\mu).
\end{equation}
Positive $\epsilon$ correspond to a particle, while negative
$\epsilon$ to a hole.

It is proven in the literature (see  Ref.~\cite{KorepinBook}) that
in case where $K(\lambda)$ is a monotonously decreasing function
of $\lambda$ for positive $\lambda$, then the solution to
\rfs{eq:exc} is a monotonously increasing function of $\lambda$
for positive $\lambda$. An example of such $K(\lambda)$ is given
by \rfs{eq:deltakernel}, or by the delta function potential.  Then
$\epsilon(\lambda)<0$ when $|\lambda|<\lambda_F$ and
$\epsilon(\lambda)>0$ when $|\lambda|>\lambda_F$. It is consistent
with the ground state where all the states with
$|\lambda|<\lambda_F$ are filled, and the rest are empty. The
excitations are then particle-like for $|\lambda|>\lambda_F$ or
hole-like if $|\lambda|<\lambda_F$.

The situation changes drastically if $K(\lambda)$ is no longer
monotonous, as in the resonant case \rfs{eq:kernel1} or
\rfs{eq:kernel2}. Then $\epsilon(\lambda)$ is no longer
necessarily a monotonous function itself.

\rfs{eq:lin} and \rfs{eq:exc} are integral equations which in
general are hard to solve analytically. In case of delta function
interactions, where $K$ is given by \rfs{eq:deltakernel}, it can
only be solved in limits of very large $c$ or very small $c$. The
corresponding solutions are described in Appendix~\ref{AppendixE}.

In case of resonant interactions, where $K$ is given by
\rfs{eq:Klarge},  we are able to find the analytic solution in two
cases: the case of moderate $c$ where we can approximate the
phaseshift by \rfs{eq:lacrude}, and in the case of small $c$ only,
where $K$ reduces to \rfs{eq:kernel1}.  We proceed to describe the
solution in these cases.

\subsection{Ground State and Excitation of Resonant Gas: the Case of
Moderate $c$}

When $c \sim \sqrt{\epsilon_0/m_a}$, it is possible to obtain the
qualitative estimate of the solution by employing the approximation
\rfs{eq:lacrude}, or \begin{equation} \label{eq:kernelcrude}
K(\lambda) \approx \frac{2 \lambda_0^2}{c \lambda_1^2}.
\end{equation} Since $K$ is a constant, solving the corresponding integral equations
is straightforward. Substitution into the Lieb-Liniger equation
\rfs{eq:lin} gives
\begin{equation} \label{eq:rhocrude}
\rho \approx \frac{1}{2\pi} \left[ 1- \frac{2 \lambda_0^2
\lambda_F}{\lambda_1^2 \pi c} \right]^{-1}.
\end{equation}
Thus the density of states is a constant, and $\lambda_F$ cannot
exceed its limit given by \rfs{eq:limit3}. We will see below, by
solving the Lieb-Liniger equation numerically, that although a crude
approximation, \rfs{eq:rhocrude} captures the main features of the
density of states: it is constant and it diverges to infinity as
$\lambda_F$ approaches its upper limit. The total density is
approximately
\begin{equation}
D=\frac{\lambda_F}{\pi}\left[ 1- \frac{2 \lambda_0^2
\lambda_F}{\lambda_1^2 \pi c} \right]^{-1}.
\end{equation}

It is equally easy to find the excitation spectrum by solving
\rfs{eq:exc}. We find
\begin{equation} \label{eq:epscrude}
\epsilon(\lambda) \approx \frac{\lambda^2}{2} -
\frac{\lambda_F^2}{2},
\end{equation}
where the chemical potential $h$ is equal to
\begin{equation}
h \approx \frac{\lambda_F^2}{6} \left( 3 - \frac{4 \lambda_0^2
\lambda_F}{c \pi \lambda_1^2} \right).
\end{equation}
Thus the spectrum remains quadratic, as for the noninteracting Fermi
gas. We can now find the Fermi velocity $v_F$ by using
\begin{equation}
v_F \approx \left. \pbyp{\epsilon}{k} \right|_{\lambda=\lambda_F}=
\frac{D}{\pbyp{D}{\lambda_F}}= \lambda_F \left[ 1- \frac{2
\lambda_0^2 \lambda_F}{\lambda_1^2 \pi c} \right]
\end{equation}
(see Ref.~\cite{KorepinBook} for derivation). Here $k$ is the
physical momentum of the excitations, given by \rfs{eq:physmom}. We
can also find the Luttinger parameter
\begin{equation}
g \approx \pi \pbyp{D}{\lambda_F} = \left[ 1- \frac{2 \lambda_0^2
\lambda_F}{\lambda_1^2 \pi c} \right]^{-2}. \end{equation} The Fermi
velocity and the Luttinger parameter both diverge as $\lambda_F$
approaches its upper limit.

\subsection{Ground State and Excitation of Resonant Gas: the Case of Small
$c$}

When $c$ is small, we can do far better than when $c$ is moderate,
and can obtain solution which is not just an estimate, but is
correct in the limit of vanishingly small $c$.

As we will see in a second, in case when $c$ is small, we only need
to consider such $\lambda_0$ that $\lambda_0 \gg \lambda_F$. Under
this condition, for the interval $|\lambda| \le \lambda_F$, we can
neglect the second and the third term of \rfs{eq:kernel1} and the
function $K$ reduces to the one for the delta function interacting
potential, given by \rfs{eq:deltakernel}, with the substitution $c_1
\rightarrow c$. The corresponding Lieb-Liniger equation, given by
\rfs{eq:LLdk}, has been studied in the literature for a long time.
In particular, its solution for $c$ being small can be found
analytically. The solution is presented in Appendix \ref{AppendixE},
and the answer for $\rho(\lambda)$ in the interval $-\lambda_F \le
\lambda \le \lambda_F$ is given by \rfs{eq:DoS}.

Now if we would like to know $\rho(\lambda)$ beyond this interval,
we can no longer neglect the second and third term of
\rfs{eq:kernel1}. However, we can use the Lieb-Liniger equation to
find $\rho(\lambda)$ for any $\lambda$ by using
\begin{equation}
\rho(\lambda) = \frac{\lambda_F}{4\pi^2 c_1}
\int_{-\lambda_F}^{\lambda_F}
d\mu~K(\lambda-\mu)\sqrt{1-\frac{\mu^2}{\lambda_F^2}}
+\frac{1}{2\pi}.
\end{equation}
Taking into account that $c_1$ and $c_2$ are small, we can
approximate $K$ by a sum of three delta functions, \begin{equation}
\label{eq:tdf} K = \delta(\lambda) + \delta(\lambda-2\lambda_0) +
\delta(\lambda+2 \lambda_0) \end{equation} to find
\begin{equation} \label{eq:rhoinside}
\rho(\lambda)  =  \frac{\lambda_F}{2 \pi c_1}
\sqrt{1-\frac{\lambda^2}{\lambda_F^2}}, \  -\lambda_F \le \lambda
\le \lambda_F,
\end{equation}
\begin{eqnarray} \label{eq:rhoright}
 &&\rho(\lambda) =\frac{\lambda_F}{2 \pi c_2}
\sqrt{1-\frac{\left(\lambda-2\lambda_0 \right)^2}{\lambda_F^2}}, \cr
&&2\lambda_0 - \lambda_F  \le \lambda  \le 2 \lambda_0 + \lambda_F,
\end{eqnarray}
and
\begin{eqnarray} \label{eq:rholeft}
&&\rho(\lambda)  =  \frac{\lambda_F}{2 \pi c_2}
\sqrt{1-\frac{\left(\lambda+2\lambda_0 \right)^2}{\lambda_F^2}}, \cr
&&-2\lambda_0 - \lambda_F \le  \lambda \le -2 \lambda_0 + \lambda_F.
\end{eqnarray}
Outside these three intervals within small $c$ approximation
$\rho(\lambda)=1/(2 \pi)$. The total density of the condensate
coincides with its delta function value, or with \rfs{eq:denstotal}
with $c$ being replaced by $c_1$, or
\begin{equation}
D=\frac{\lambda_F^2}{4 c_1}.
\end{equation}

Now we use the same technique to find $\epsilon(\lambda)$. Within
the interval $-\lambda_F\le \lambda\le \lambda_F$ the solution is
simply given by the corresponding delta-function interaction
solution, \rfs{eq:sp1}. Outside this interval, we can use the
approximation \rfs{eq:tdf}. Thus we find
\begin{equation} \label{eq:epsinside}
\epsilon(\lambda) = - \frac{\lambda_F^3}{6 c_1} \left( 1-
\frac{\lambda^2}{\lambda_F^2} \right)^{\frac{3}{2}}, \ |\lambda| \le
\lambda_F,
\end{equation}
\begin{eqnarray} \label{eq:epsright}
&&\epsilon(\lambda) =\frac{\lambda^2}{2}-\frac{\lambda_F^2}{4}-
\frac{\lambda_F^3}{6c_1} \left( 1-\frac{\left(\lambda-2 \lambda_0
\right)^2}{\lambda_F^2} \right)^{\frac{3}{2}}, \cr && 2 \lambda_0 -
\lambda_F \le \lambda \le 2 \lambda_0 + \lambda_F,
\end{eqnarray}
and
\begin{eqnarray} \label{eq:epsleft}
&&\epsilon(\lambda) =\frac{\lambda^2}{2}-\frac{\lambda_F^2}{4}-
\frac{\lambda_F^3}{6c_1} \left( 1-\frac{\left(\lambda+2 \lambda_0
\right)^2}{\lambda_F^2} \right)^{\frac{3}{2}}, \cr && -2 \lambda_0 -
\lambda_F \le \lambda \le -2 \lambda_0 + \lambda_F.
\end{eqnarray}
Outside either of the three intervals \begin{equation}
\label{eq:epsoutside} \epsilon(\lambda) =
\frac{\lambda^2}{2}-\frac{\lambda_F^2}{4}.
\end{equation}
We emphasize that these formulae break down around the ends of the
three appropriate intervals. We will see in the next section where
we solve the Lieb-Liniger equation numerically, that they give a
rather good approximation to the actual solution.

Looking at the expressions for $\epsilon(\lambda)$ we indeed find
that it is not a monotonous function of $\lambda$. In fact, it has a
minimum approximately at $\lambda \approx \pm 2 \lambda_0$. The
minimum occurs around momenta whose particles are in resonance with
Fermi sea. This is because two particles with momenta $2 \lambda_0$
and $0$ have a relative momentum $\lambda_0$ in their center of mass
frame. Thus they are in resonance.

Moreover, it is easy to convince oneself that if $\lambda_F$ is
increased at a fixed $\lambda_0$, at a certain critical value of
$\lambda_F$  the minimum becomes negative and the excitation
spectrum becomes unstable. The critical value of the Fermi momentum
when the minimum crosses zero can be also approximately found by
equating \rfs{eq:epsright} with zero and is given by
\begin{equation} \label{eq:lambda0crit}
\lambda_F=\left(12 c_1 \lambda_0^2 \right)^\frac{1}{3}.
\end{equation}
Upon substituting for $c_1$ and expressing $\lambda_1$, $\lambda_0$
in terms of the parameters of the Hamiltonian, we find
\rfs{eq:lcrit0}. Notice that this critical value of $\lambda_F$ is
much less than $\lambda_0$, thus the approximation used to derive
these results indeed holds. This leads to the critical value of
density
\begin{equation} \label{Dcrit}
D_{\rm crit}= \left( \frac{ 9 \lambda_0^4 }{4 c_1} \right)^\ot.
\end{equation}
This is equivalent to \rfs{eq:Dcrit0}.  Above this critical density
the Bose gas with resonant interactions as in \rfs{eq:FA} with $c$
small no longer has a simple ground state described in this section
of the paper, as the function $\epsilon(\lambda)$ becomes negative
at momenta higher than $\lambda_F$.

To check these results, we are going to compare them with the
numerical solution of the Lieb-Liniger and excitation spectrum
equations.

\subsection{Numerical Solution}

For the purpose of numerics, it is advantageous to pass to
dimensionless variables by the appropriate rescaling
\begin{equation} \label{eq:rescaling}
\lambda=\lambda_F \t \lambda, \ \mu = \lambda_F \t \mu, \
c=\lambda_F \t c, \epsilon=\lambda_F^2 \t \epsilon, \  h =
\lambda_F^2 \t h.
\end{equation}
$\rho$ does not need to be rescaled. Then the Lieb-Liniger and the
energy spectrum equations become
\begin{equation} \label{eq:linq}
\rho(\t \lambda)-\frac{1}{2 \pi} \int_{-1}^{1} d \t \mu~\t K(\t
\lambda-\t \mu) \rho(\t \mu) = \frac{1}{2 \pi},
\end{equation}
and
\begin{equation} \label{eq:excq}
\t \epsilon(\t \lambda)-\frac{1}{2\pi} \int_{-1}^{1} d\t \mu~\t
K(\t \lambda-\t \mu) \t \epsilon(\t \mu) = \frac{\t
\lambda^2}{2}-\t h.
\end{equation}
Here $\t K$ denotes the functions given in \rfs{eq:kernel1} or
\rfs{eq:kernel2}, but with $\t c_1=c_1/\lambda_F$, $\t
c_2=c_2/\lambda_F$, $\t c_3=c_3/\lambda_F$, $\t
\lambda_0=\lambda_0/\lambda_F$, and $\t
\lambda_1=\lambda_1/\lambda_F$ replacing the variables $c_1, c_2,
c_3, \lambda_0, \lambda_1$ in them.

These dimensionless equations have been solved numerically. We used
a simple discretization procedure where the interval between $-1$
and $1$ was divided into 1200 intervals, and the integrals in the
equations were computed via a straightforward Newton discretization.
The resulting system of linear  equations was solved by inverting
the kernel matrix.

\subsubsection{Numerical solution for moderate $c$}
We take $\lambda_0=2$, $\lambda_1=1$, $c=5$. Under these conditions,
the upper limit \rfs{eq:limit3} for $\lambda_F$ is around $2$. We
solve the integral equations \rfs{eq:lin} and \rfs{eq:exc} for
$\lambda_F=1.96$, close to its upper limit.
\begin{figure}[tb]
\includegraphics[height=1.5in]{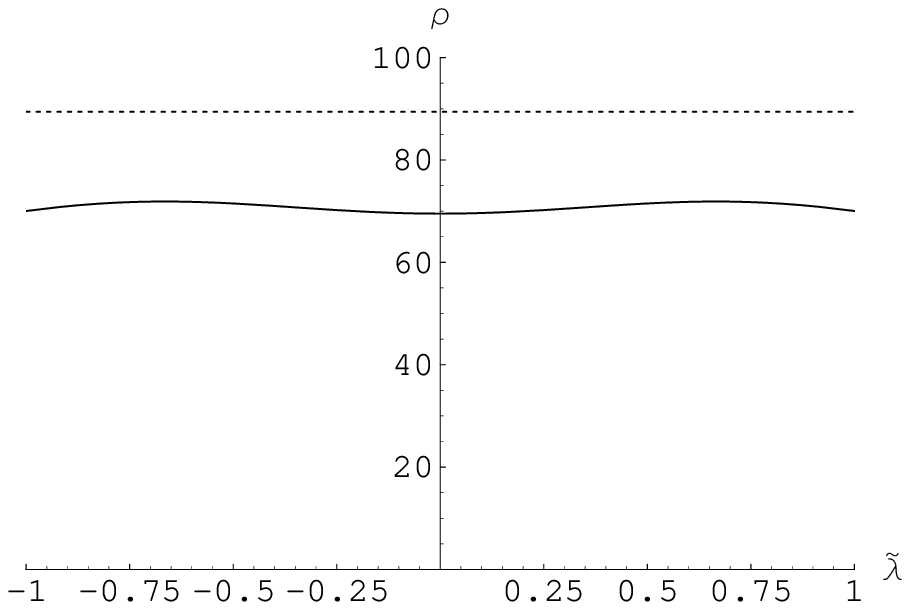}
\caption{\label{Fig:rho1crude} $\rho(\t \lambda)$ as a function of
$\t \lambda$. Here $K$ is taken from \rfs{eq:Klarge} with $\t
\lambda_0=2$, $\t \lambda_1=1$, and $\t c=5$. Dashed curve is the
analytic estimate \rfs{eq:rhocrude}.}
\end{figure}
Fig.~\ref{Fig:rho1crude} shows the density of states (solid line)
and the analytic estimate \rfs{eq:rhocrude} (dashed line). Although
not coinciding, the estimate captures the fact that $\rho$ is close
to a constant, and that it diverges as $\lambda_F$ approaches its
upper limit.

\begin{figure}[tb]
\includegraphics[height=1.5in]{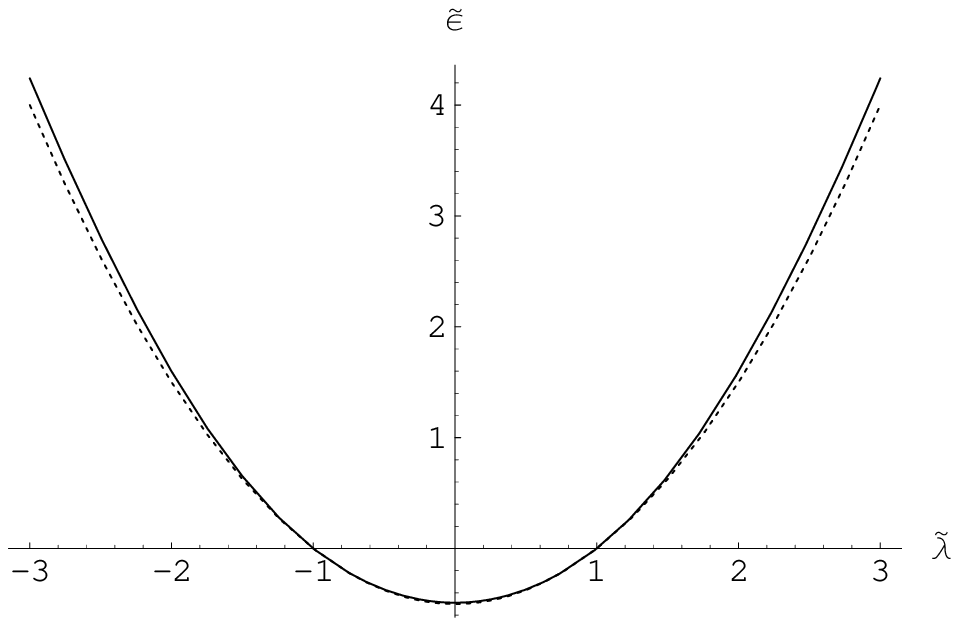}
\caption{\label{Fig:eps1crude} $\t \epsilon(\t \lambda)$ as a
function of $\t \lambda$. Here $K$ is taken from \rfs{eq:Klarge}
with $\t \lambda_0=2$, $\t \lambda_1=1$, and $\t c=5$. Dashed curve
is the analytic estimate \rfs{eq:epscrude}.}
\end{figure}

Fig.~\ref{Fig:eps1crude} shows the excitation spectrum, both
numerical and the analytic estimate \rfs{eq:epscrude} (shown as a
dashed line).

\subsubsection{Numerical solution for small $c$}
 We take $K$ from \rfs{eq:kernel1}. First, we consider
the case when $\t \lambda_0 \gg 1$, or in other words the energy of
the resonance $\lambda_0$ is much bigger than the Fermi momentum
$\lambda_F$.

\begin{figure}[tb]
\includegraphics[height=1.5in]{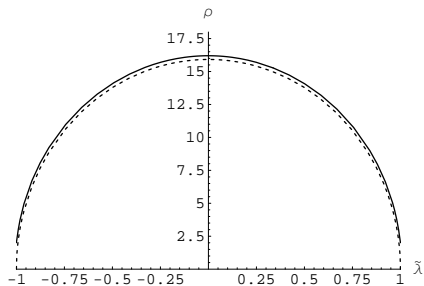}
\caption{\label{Fig:rho1} $\rho(\t \lambda)$ as a function of $\t
\lambda$. Here $K$ is taken from \rfs{eq:kernel1} with $\t c_1=\t
c_2=0.01$, and $\t \lambda_0=10$. Solid line represents numerics,
while the dashed line is the analytic prediction.}
\end{figure}

Fig.~\ref{Fig:rho1} presents the graph of $\rho(\t \lambda)$
obtained numerically for the case when $\t \lambda_0=10$ and $\t
c_1=\t c_2=0.01$.
For comparison, the analytic prediction \rfs{eq:rhoinside} is
plotted as a dashed line which generally lies slightly below the
numerical line. This graph is indeed very close to
\rfs{eq:rhoinside}, except at $\t \lambda=\pm 1$ where, as discussed
in the Appendix~\ref{AppendixE}, the analytic solution breaks down.

\begin{figure}[tb]
\includegraphics[height=1.5in]{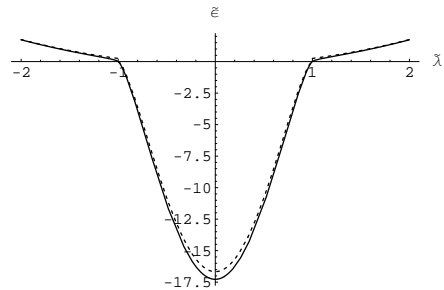}
\caption{\label{Fig:eps1} $\t \epsilon(\t \lambda)$. Here $K$ is
taken from \rfs{eq:kernel1} with $\t c_1=\t c_2=0.01$, and $\t
\lambda_0=10$. The solid curve represents numerics, and the dashed
line is the analytic prediction. }
\end{figure}

Fig.~\ref{Fig:eps1} presents the graph of $\t \epsilon(\t \lambda)$.
The analytic predictions, which follow from \rfs{eq:epsinside} and
\rfs{eq:epsoutside}, also work rather well. In fact, for
$|\lambda|>\lambda_F$, we used \rfs{eq:epsoutside} as opposed to a
more precise \rfs{eq:sp2}. Had we used this latter expression, the
analytic curve would have been completely indistinguishable from the
numerical line.

\begin{figure}[tb]
\includegraphics[height=1.5in]{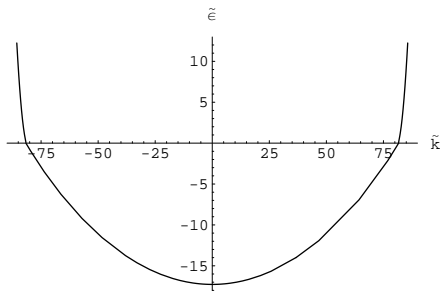}
\caption{\label{Fig:eps1a} $\t \epsilon(\t k)$ as a function of the
momentum of the excitations $\t k$, computed numerically. Here $K$
is taken from \rfs{eq:kernel1} with $\t c_1=\t c_2=0.01$, and $\t
\lambda_0=10$. }
\end{figure}

Finally, Fig.~\ref{Fig:eps1a} presents the graph of $\t \epsilon(\t
k)$, with $\t k$ being the rescaled momentum of the excitations, $\t
k = k/\lambda_F$, where $k$ is given by \rfs{eq:physmom}. For
simplicity, we approximated $\delta$ in \rfs{eq:physmom} by $\delta
\approx \pi  \sign \,(\lambda)+\pi \sign \, (\lambda-2\lambda_0)
+\pi \sign \,(\lambda+2 \lambda_0)$.
This graph also closely follows the analytic prediction given by
Eqs.~\rf{eq:sp1}, \rf{eq:sp2}, and \rf{eq:momnar}. We would like to
remark that this graph shows that the excitation spectrum at
$\lambda \approx \pm \lambda_F$ undergoes a sharp change in the
slope. It is interesting that the Luttinger liquid theory
concentrates on a tiny interval around the Fermi points where the
spectrum can be linearized. It is clear that this theory cannot tell
us anything about the spectrum away from these points (the interval
where Luttinger liquid theory applies is of course so small because
of the small value of $\t c$).

We expect that as the density of the gas is increased, or in other
words, as $\t \lambda_0$ is decreased, the excitation spectrum will
develop minima. At the critical value of $\t \lambda_0$, given by
\rfs{eq:lambda0crit}, the minimum of the spectrum will dip below
zero. At $\t c_1=0.01$, this value is $\t \lambda_0 \approx 2.89$.
We solve the integral equations numerically at $\t \lambda_0=3.3$.

\begin{figure}[tb]
\includegraphics[height=1.5in]{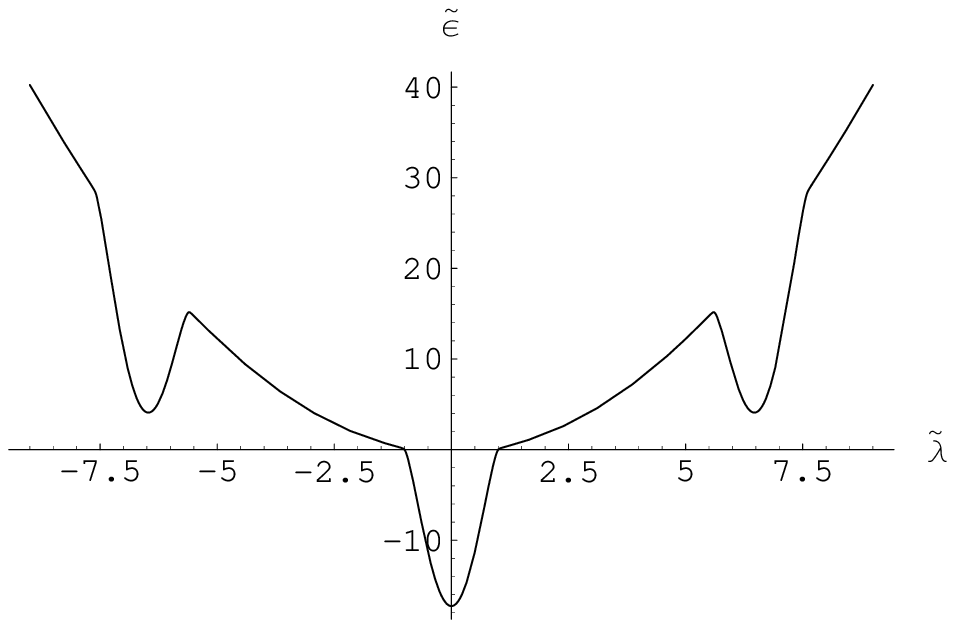}
\caption{\label{Fig:eps2a} $\t \epsilon(\t \lambda)$ as a function
of $\t \lambda$. Here $K$ is taken from \rfs{eq:kernel1} with $ \t
c_1=\t c_2=0.01$, and $\t \lambda_0=3.3$. }
\end{figure}

\begin{figure}[tb]
\includegraphics[height=1.5in]{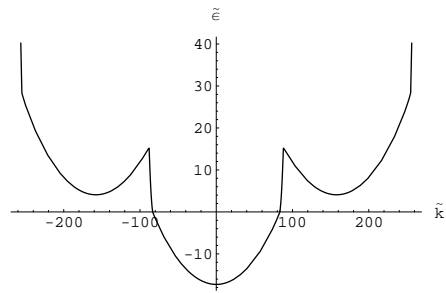}
\caption{\label{Fig:eps2} $\t \epsilon(\t k)$ as a function of the
momentum of the excitations $\t k$. Here $K$ is taken from
\rfs{eq:kernel1} with $\t c_1=\t c_2=0.01$, and $\t \lambda_0=3.3$.
}
\end{figure}

The density of states in the interval $-1 \le \t \lambda \le 1$
remains basically unchanged, as predicted. Let us now look at the
excitation spectrum. We present here both the spectrum as a function
of $\t \lambda$, and as a function of $\t k$, see
Fig.~\ref{Fig:eps2a} and \ref{Fig:eps2}. It differs drastically from
that on Fig.~\ref{Fig:eps1}.
Namely it developed an additional minimum at $\lambda \approx 2
\lambda_0$. We only plot numerical curves on these graphs as the
analytic curves given by Eqs.~\rf{eq:epsinside}, \rf{eq:epsright},
\rf{eq:epsleft}, and \rf{eq:epsoutside} are almost indistinguishable
from the numerical ones.

We can understand the appearance of a minimum at $\lambda=2
\lambda_0$ by noticing that the particles which move at
$\lambda=2\lambda_0$ scatter off the bulk of the Fermi sea exactly
at resonance. Therefore, these particles try to form bound states
with those in the Fermi sea and the resulting bosons then Bose
condense.

\begin{figure}[tb]
\includegraphics[height=1.5in]{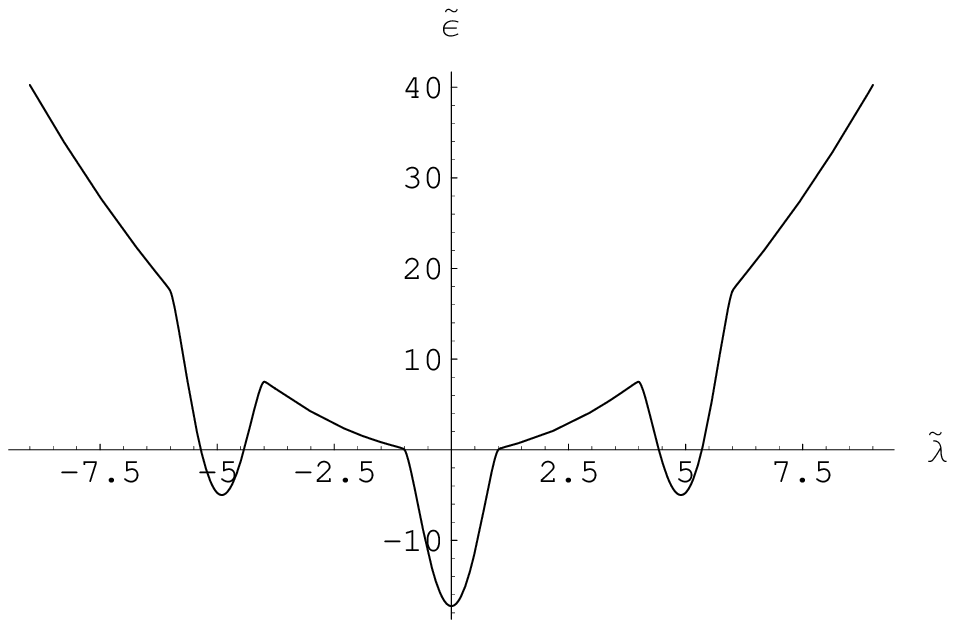}
\caption{\label{Fig:eps3a} $\t \epsilon(\t \lambda)$ as a function
of $\t \lambda$. Here $K$ is taken from \rfs{eq:kernel1} with $\t
c_1=\t c_2=0.01$, and $\t \lambda_0=2.5$. }
\end{figure}

\begin{figure}[tb]
\includegraphics[height=1.5in]{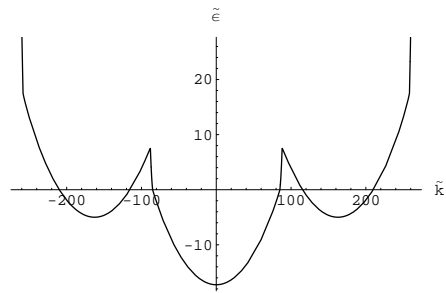}
\caption{\label{Fig:eps3} $\t \epsilon(\t k)$ as a function of the
momentum of the excitations $\t k$. Here $K$ is taken from
\rfs{eq:kernel1} with $\t c_1=\t c_2=0.01$, and $\t \lambda_0=2.5$.
}
\end{figure}

Now let us look the spectrum at $\t \lambda_0=2.5$. The density of
states is again not much different from that on Fig.~\ref{Fig:rho1},
but the spectrum changes again, and is shown on Fig.~\ref{Fig:eps3a}
and Fig.~\ref{Fig:eps3}.
Remarkably, the spectrum develops areas where $\epsilon<0$. By
filling up the states where $\epsilon$ is negative, it is possible
to lower the energy of the system compared to that of the ground
state. Therefore, the state discussed in this section becomes
unstable, and is no longer the ground state in this regime. It can
be checked numerically that this instability develops roughly at
$\t\lambda_0 \approx 2.9$, just as predicted analytically. This is
why we refer to the state where only the states between $-\lambda_F$
and $\lambda_F$ are filled as ``simple ground state". At
sufficiently high density of the gas the ground state changes.

\subsubsection{Numerical solution at large $c$}

If $c$ is very large, then the kernel $K$ can be taken from
\rfs{eq:kernel2}. We are not able to find an analytical solution for
this case, and present the numerical solution. It is qualitatively
similar to the one at small $c$, although it has several new
features.

We take $\t c_3 =0.01$, and present numerical solution for $\t
\lambda_1=0.65$, above the instability, and for $\t \lambda_1=0.55$,
below the instability. $\t \lambda_1$ can generally take values
smaller than appropriate values of $\lambda_0$ at small $c$ because
of the condition \rfs{eq:limit2} as compared to \rfs{eq:limit1}.
\begin{figure}[tb]
\includegraphics[height=1.5in]{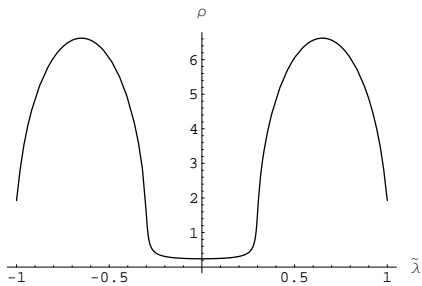}
\caption{\label{Fig:eps4} The density of states $\rho(\t \lambda)$
for the case of large $c$, \rfs{eq:kernel2}. $\t c_3=0.01$, $\t
\lambda_1 = 0.65$. }
\end{figure}
The density of states at $\lambda_1=0.65$ is shown on
Fig.~\ref{Fig:eps4}. Notice it is quite different from the
analytically understood case of small $c$, depicted on
Fig.~\ref{Fig:rho1}.

\begin{figure}[tb]
\includegraphics[height=1.5in]{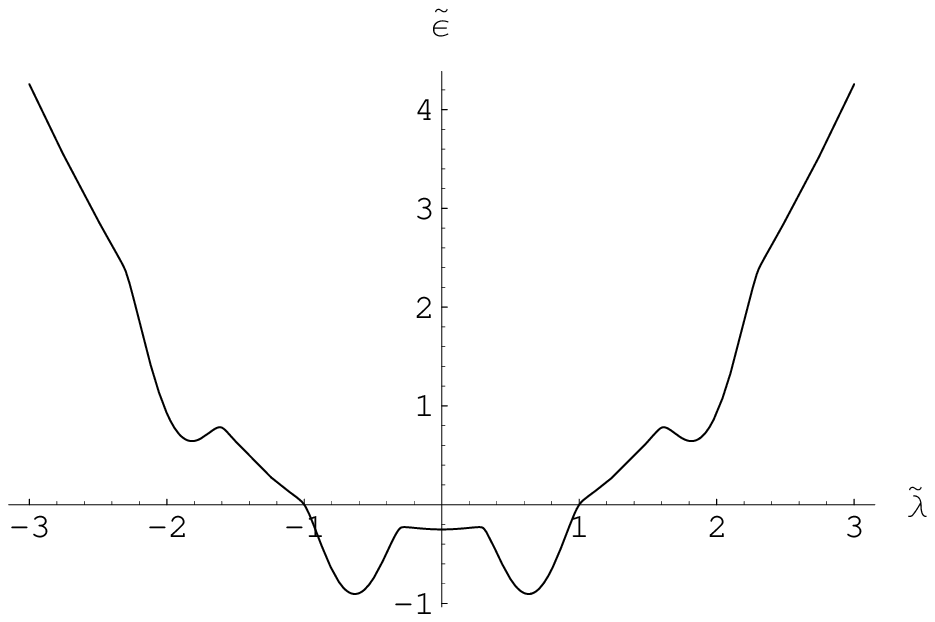}
\caption{\label{Fig:eps4a} $\t \epsilon(\t \lambda)$ as a function
of $\t \lambda$. Here $K$ is taken from \rfs{eq:kernel2} with
$c_3=0.01$, and $\lambda_1=0.65$. }
\end{figure}

As far as the energy spectrum, it is depicted on
Fig.~\ref{Fig:eps4a}. It has minima similar to the ones on
Fig.~\ref{Fig:eps1a}. In addition, not only particles, but also
holes have nonmonotonous spectrum.
\begin{figure}[tb]
\includegraphics[height=1.5in]{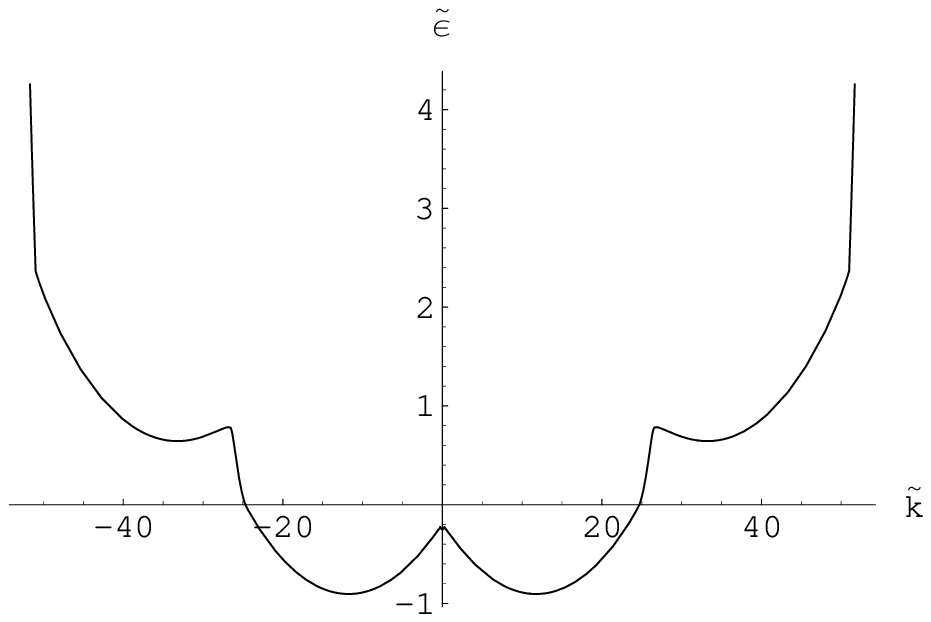}
\caption{\label{Fig:eps4b} $\t \epsilon(\t k)$ as a function of the
physical momentum $\t k$. Here $K$ is taken from \rfs{eq:kernel2}
with $\t c_3=0.01$, and $\t \lambda_1=0.65$. }
\end{figure}
Fig.~\ref{Fig:eps4b} shows the same spectrum, but as a function of a
physical momentum $k$.

\begin{figure}[tb]
\includegraphics[height=1.5in]{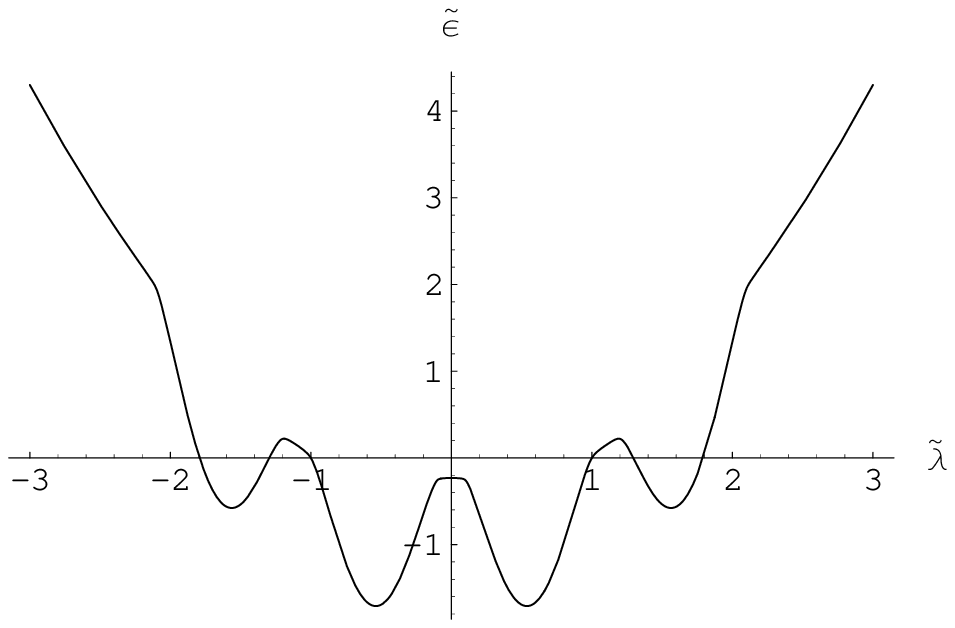}
\caption{\label{Fig:eps5a} $\t \epsilon(\t \lambda)$ as a function
of $\t \lambda$. Here $K$ is taken from \rfs{eq:kernel2} with $\t
c_3=0.01$, and $\t \lambda_1=0.55$. }
\end{figure}

\begin{figure}[tb]
\includegraphics[height=1.5in]{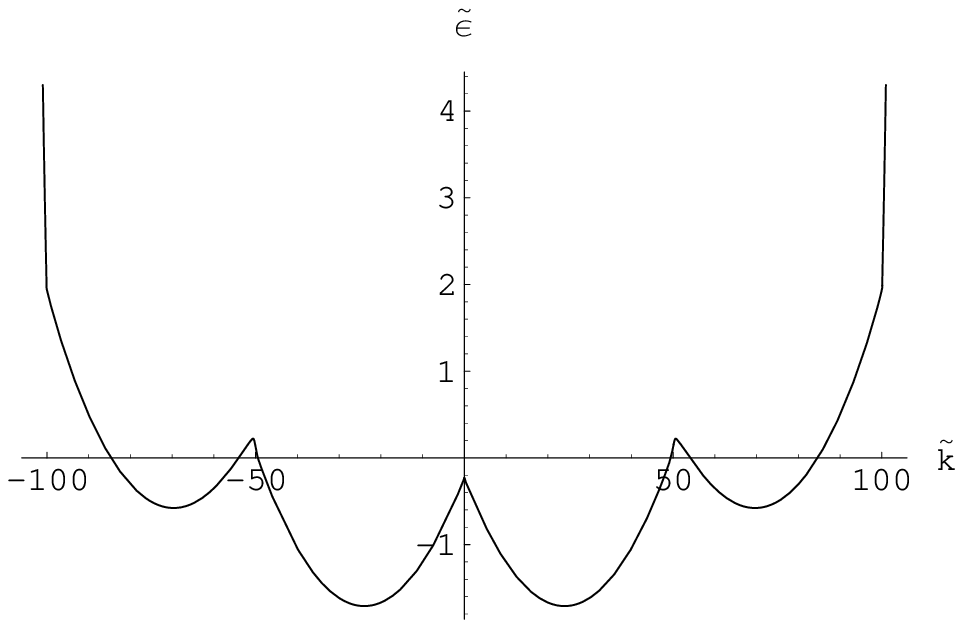}
\caption{\label{Fig:eps5b} $\t \epsilon(\t k)$ as a function of the
physical momentum $\t k$. Here $K$ is taken from \rfs{eq:kernel2}
with $\t c_3=0.01$, and  $\t \lambda_1=0.65$. }
\end{figure}

Fig.~\ref{Fig:eps5a} and \ref{Fig:eps5b} show the excitation
spectrum for $\lambda_0 = 0.55$. Now the spectrum is unstable.

\section{Beyond the simple ground state}
\label{sec:BSGS}

If the density of gas is increased beyond the critical density, at
small $c$ or at large $c$, then we can no longer use the
Lieb-Liniger and excitation integral equations Eqs.~\rf{eq:lin} and
\rf{eq:exc}. Instead, we can take a different path.

The problem we confront is that above the critical density we no
longer know which states are filled and which states are empty in
the ground states. We can circumvent this problem by writing down
the zero temperature version of the Yang-Yang equation
\cite{KorepinBook}, which looks like
\begin{equation}
\label{eq:YY} \epsilon(\lambda) -
\frac{1}{4\pi}\int_{-\infty}^\infty d\mu~K(\lambda-\mu)
\left[\epsilon(\mu) - \left|\epsilon(\mu) \right| \right] =
\frac{\lambda^2}{2}-h.
\end{equation}
This equation has to be solved at a given chemical potential $h$.
Once its solution is known, we know that the particles states are
those where $\epsilon(\lambda)>0$, while the hole states are where
$\epsilon(\lambda)<0$. We can now determine the density of states by
solving \begin{equation} \rho(\lambda) -
\frac{1}{4\pi}\int_{-\infty}^\infty d\mu ~K(\lambda-\mu) \rho(\mu)
\left[1-\sign~\epsilon(\lambda) \right]=\frac{1}{2 \pi}.
\end{equation}
In case of low density resonant gas it is easy to solve \rfs{eq:YY}
as the integration there is effectively over the region where
$\epsilon(\lambda)$ is negative. In those simple cases, we know that
$\epsilon(\lambda)$ is negative only for $-\lambda_F \le \lambda\le
\lambda_F$ and \rfs{eq:YY} reduces to \rfs{eq:exc}. But above the
critical density, it is not at all clear how to solve these
equations, either analytically or numerically. Numerically, any
naive scheme one could use to solve \rfs{eq:YY} becomes unstable.
For example, trying to find the solution  by neglecting the integral
in \rfs{eq:YY} first, and then taking it into account via
iterations, immediately leads to an instability and a solution which
quickly diverges with each successive iteration. In fact, it is
proven in Ref.~\cite{KorepinBook} that solving \rfs{eq:YY} by
iterations works. However, their proof essentially uses the
monotonicity of $K(\lambda)$ for $\lambda>0$, that is, it breaks
down for resonant interactions.

It is intuitively clear that above the critical density the single
particle states whose momentum $\lambda$ falls into the areas of
negative $\epsilon(\lambda)$, such as on Fig.~\ref{Fig:eps3a}, will
be filled. But all of those states cannot be filled. Otherwise, the
analog of the condition \rfs{eq:conditionnec} (which is the very
same condition, except the regions of integration are now over where
$\epsilon(\lambda)$ is negative) would be violated. Unfortunately,
it is not clear how to improve this intuition and make it more
quantitative.

Solving \rfs{eq:YY} and studying the resonant gas above the critical
density will be left for future work.

\section{Conclusions and Outlook}
We presented the technique for solving the problem of one
dimensional resonantly interacting Bose gas exactly. With the
Hamiltonian given by \rfs{eq:FA}, supplemented by the condition
\rfs{eq:nbs}, we found the solution analytically for small $c$, or
$c \ll \sqrt{\epsilon_0/m_a}$, at density below the critical density
\rfs{eq:Dcrit0}. In particular, we found the exact excitation
spectrum, given by Eqs.~\rf{eq:epsinside}, \rf{eq:epsright},
\rf{eq:epsleft}, and \rf{eq:epsoutside}. The excitation spectrum is
not monotonous and has minima above the Fermi momentum. A typical
excitations spectrum, as a function of asymptotic momentum $\lambda$
is shown on Fig.~\ref{Fig:eps2a}. An excitation spectrum as a
function of a physical momentum of a quasiparticle is shown on
Fig.~\ref{Fig:eps2}. Above the critical density the solution we
found becomes unstable and we were not able to find a stable
solution in this regime. These features of the solution to the
resonantly interacting gas remain valid at large  $c$ as well, as
confirmed by numerical solutions of the Bethe Ansatz equations.

The minima in the excitation spectrum should be easily observable
experimentally, if such a one dimensional gas is realized in a
laboratory. Indeed, at any finite temperature, one should observe
low energy excitations far above the Fermi momentum. These could be
observed in a time of flight experiment.

At $c \sim \sqrt{\epsilon_0/m_a}$, the excitation spectrum does not
develop minima and the gas can be described by the Luttinger liquid
theory. However, in one difference from the nonresonantly
interacting Bose gas, its Fermi momentum $\lambda_F$ can never
exceed a certain cutoff momentum in this regime. The increased
density of gas is accommodated by the ever increasing density of
states, as opposed to the Fermi momentum.

In view of the possible future experiments, it is important to
generalize the theory presented here to nonzero temperature. This
would be straightforward to do by employing the appropriate
Yang-Yang equation.

The nature of the phase above the critical density for large or
small $c$ remains undetermined, and should be another subject of
future work.

A major limitation of the technique used in this paper is the
neglect of three-body processes. Although justified in the
relatively low density regimes considered here, a major technical
improvement would involve coming up with a technique where this
deficiency is absent. Perhaps an integrable version of \rfs{eq:FA}
can be formulated or some other technique can be proposed. Whether
this is possible to do is currently not clear. In the absence of
exact techniques, bosonization remains the only alternative if one
wants to study high density regimes where the density of molecules
is supposed to be of the order of the density of atoms.

Finally, an interesting generalization of the work presented here
would be to study a gas of fermions resonantly interacting and
confined to one dimension. The main difficulty which will have to be
overcome is that fermions carry spin, the $S$-matrix becomes a real
matrix in the spin space, and the Bethe Ansatz techniques become
much more technically involved than in the case considered in this
paper.

\acknowledgements The author is grateful to A. Tsvelik and R. Konik
for useful comments at the beginning of this work, and to L.
Radzihovsky and D. Sheehy for many discussions. This work was
supported by the NSF grant DMR-0449521.

\appendix
\section{Low Energy One Dimensional Scattering}
\label{AppendixA} We would like to derive formulae
\rfs{eq:invform}. Closely following Ref.~\cite{LL}, we observe
that
$$
|S_{S,A}|^2=|1+2 f_{S,A}|^2=1.
$$
It follows from here that
$$
2+\frac{1}{f_{S,A}}+\frac{1}{f^*_{S,A}}=0,
$$
and thus, Re $f_{S,A}^{-1}=-1$. We deduce
$$
f_{S,A}=\frac{1}{i \t F_{S,A}(k)-1},
$$
where $\t F_{S,A}$ is a function of momentum $k$, real when $k$ is
real. Notice that
$$\cot \delta_{S,A} = -\t F_{S,A}(k).$$

Next we would like to analyze the behavior of $\t F$ at very small
momenta. For that we need to study the Schr\"odinger equation at
small energy. We observe that it is possible to split the range of
$x$ into several intervals. In the interval $-r_0 \lesssim x
\lesssim r_0$, we can neglect energy to find the Schr\"odinger
equation
\begin{equation} \label{eq:wse}
-\frac{1}{2m} \frac{\partial^2 \psi}{\partial x^2}+U(x)\psi=0,
|x|<r_0
\end{equation}
This equation must be solved with the boundary conditions
$\psi(0)=1$, $\dbyd{\psi(0)}{x}=0$ for symmetric amplitude, and
with the conditions $\psi(0)=0$, $\dbyd{\psi(0)}{x}=1$ if we study
antisymmetric amplitude.

In the interval $r_0 \lesssim |x| \lesssim 1/k$, we neglect the
potential and the energy to arrive at
$$
-\ohm \frac{\partial^2 \psi}{\partial x^2} =0.
$$
Its solution is $\psi=c_1+c_2 x$. The constants $c_1$ and $c_2$
can be determined by matching this solution with the solution to
\rfs{eq:wse} in the range of $|x| \sim r_0$. The crucial point is
that they are energy independent. They of course depend on whether
we are looking at symmetric or antisymmetric amplitude.

Finally in the range $|x| \gtrsim 1/k$, the potential (but not the
energy) can be neglected, and the Schr\"odinger equation reduces
to
$$
-\ohm \frac{\partial^2 \psi}{\partial x^2} = \frac{k^2}{2m} \psi.
$$
The solution to this reads $A \cos(kx+\delta_S)$ in the symmetric
case or $A  \sin(kx+\delta_A)$ in the antisymmetric case, if $x
\gtrsim 1/k$ and their symmetric (antisymmetric) reflection if $x
\lesssim -1/k$. Extending it to $|x| < 1/k$, we find $\cot
\delta_S=-k c_1/c_2$. Thus $\t F_S=k c1/c2$ at small $k$, and we
arrive at
$$ f_S=\frac{1}{i k F_S(k^2)-1},$$
where $F_S(k)=\t F_S(k)/k$, and $F_S(0)$ is finite. Quite
analogously, taking the asymptotics $A \sin(kx+\delta_A)$, we find
$k \cot \delta_A=c_2/c_1$, leading to
$$ f_A=\frac{1}{\frac{i}{k} F_A(k^2)-1},$$
where $F_A(0)$ is finite.

Finally, we observe that for purely imaginary $k$ the scattering
amplitude must be real. For that the functions $F_S$ and $F_A$
must be even functions of momentum $k$, which explains why they
are the functions of $k^2$.

\section{$T$-matrix calculation of the scattering amplitude}
\label{AppendixB}

We would like to calculate the $T$-matrix of the $a$-particles
described by \rfs{eq:Fano}. The definition of $T$-matrix is given in
many textbook, see, for example, Ref.~\cite{HewsonBook}. It is
basically the exact Green's function with free Green's function
subtracted and external legs amputated.

We note that the free Green's function is given by
$$G_0(E;k)=\frac{1}{E-\frac{k^2}{2m}+i 0},$$
and the Green's function of $b$-particles is
$$D(E)=\frac{1}{E-\epsilon_0+i 0}.$$
The $T$-matrix is given by the geometric
series
$$T=\t c+\t c \t G \t c+ \dots=\frac{\t c}{1-\t G \t c}.$$
Here $\t G$ represents trace of the Green's function
$$ \t G = \int_{\infty}^\infty \frac{dk}{2\pi} \frac{1}{E-\frac{k^2}{2m}+i
0}=-i\sqrt{\frac{m}{2  E}},$$ and $\t c$ is
$$ \t c = c+gDg.$$
Thus the $T$-matrix is
$$ T=\frac{c+gDg}{1-\t G \left(c+gDg \right)}=\left( \frac{E-\epsilon_0}{
c\left(E-\epsilon_0 \right)+g^2}-\t G \right)^{-1}.$$ Upon
substituting $E=k^2/2m$ and using \rfs{eq:ts} we get \rfs{eq:sc}.

\section{Bethe Ansatz for the many-body wave function}
\label{AppendixC} We would like to discuss why \rfs{eq:sol} solves
the appropriate Schr\"odinger equation. The proof is known from the
studies of Bethe Ansatz. We are not going to present a mathematical
proof, but merely list the reasons why this is the case.

If $x_1 \not = x_2 \not  = \dots \not = x_n$, then clearly this wave
function solves the Schr\"odinger equation by construction, with the
energy given by \rfs{eq:energy}.

Let us now take $x_j$ close to $x_{k}$ so that there are no other
particles between these two. The terms in \rfs{eq:sol} can be split
into pairs which differ by exchanging $x_j$ and $x_{k}$. Let's take
one such pair. Then if $x_j < x_{k}$, the wave function which
includes the sum over this pair will be proportional to
$$\psi \sim e^{i \lambda_1 x_j+i \lambda_{2} x_{k}}+e^{i \lambda_1
x_{k}+i \lambda_{2} x_{j}} S\left( \frac{\lambda_1-\lambda_2}{2}
\right),$$ where $\lambda_1$ and $\lambda_2$ are two momenta which
$x_j$ and $x_{k}$ get multiplied by in the exponential in each of
this pair of terms. If, however, $x_j>x_{k}$, then the wave function
is proportional to
$$ \psi \sim e^{i \lambda_1 x_j + i \lambda_2 x_{k} } S\left(
\frac{ \lambda_1-\lambda_2}{2} \right)+ e^{i \lambda_2 x_j + i
\lambda_1 x_{k}}.$$ Introducing relative coordinates
$$R=\frac{x_j+x_{k}}{2}, \ r = x_j-x_{k},$$
$$\lambda=\frac{\lambda_1-\lambda_2}{2}, \
\Lambda=\lambda_1+\lambda_2,$$ we find
$$ \psi \sim e^{\Lambda R} \left( e^{i \lambda r} + S(\lambda) e^{-i
\lambda r}\right),$$ for $r<0$ and
$$ \psi \sim e^{\Lambda R} \left( e^{-i \lambda r} + S(\lambda) e^{i
\lambda r}\right),$$ for $r>0$. This coincides with the definition
of the scattering matrix \rfs{eq:sw}. So \rfs{eq:sol} indeed solves
the Schr\"odinger equation even when the particles pass through each
other.

Now let us show that if $\lambda_j=\lambda_k=\lambda$, then $\t
\psi$, defined in \rfs{eq:perm}, is antisymmetric under the exchange
of $x_j$ and $x_k$. Indeed, suppose that $x_j < x_k$. Then once
$x_j$ and $x_k$ are exchanged, we need to bring $x_k$ through all
the particles separating $x_j$ and $x_k$ until it occupies the
position $x_j$ normally occupies. At the same time $x_j$ should be
brought back to the position it normally occupies. Each exchange of
$x_k$ with the particles $x_l$ in between brings
$S\left(\frac{\lambda-\lambda_l}{2}\right)$, where $l$ labels these
particles. Each exchange of $x_j$ brings the factor $S\left(
\frac{\lambda_l-\lambda}{2} \right)$. Finally, an exchange of $x_j$
and $x_k$ brings the factor $S(0)$. Thus $\t \psi$, where $x_j$ and
$x_k$ were exchanged, differs from $\t \psi$ before the exchange, by
the factor of
\begin{equation}
S(0) \prod_l S\left(\frac{\lambda-\lambda_l}{2}\right)
S\left(\frac{\lambda_l-\lambda}{2} \right)=-1.
\end{equation}
Here the product is over all the particles between $x_j$ and $x_k$.
And indeed, $S(\lambda) S(-\lambda)=1$, as follows directly from the
form
$$S(\lambda)=\frac{i \lambda F(\lambda^2)+1}{i \lambda
F(\lambda^2)-1},$$ which follows from \rfs{eq:invform}. At the same
time $S(0)=-1$, as follows from the same relation.

This important property, from which the bosonic ``Pauli principle"
immediately follows, has one notable exception. Take the scattering
amplitude \rfs{eq:basicS} and set $\epsilon_1=0$. This corresponds
to $\epsilon_0-g^2/c=0$, or in other word, to the situation where
there is a bound state exactly at zero energy. Then, as can be
easily checked, $S(0)=1.$ Thus in these circumstances there is no
bosonic ``Pauli principle". Fortunately, this only happens when the
condition of interest to us, \rfs{eq:nbs}, is violated.

\section{Ground state and excitations  of the Bose gas with delta function interactions}
\label{AppendixE}

Consider the Bose gas \rfs{eq:mbs} with repulsive two body
potential $U(x)=c~\delta(x)$ ($c>0$), and with total density
(number of particles per unit length) $D$.

The solution to this problem can be found by solving the equations
of Bethe Ansatz. Generally this involves solving integral
equations which can only be solved numerically. The limits of very
strong or very weak interactions can nevertheless be solved
analytically. We present the solutions here.

If the interaction is very strong ($c\gg D$), then the kernel of
the integral equation, \rfs{eq:deltakernel} vanishes. The
Lieb-Liniger equation can be solved elementary, to give
$$
\rho(\lambda) = {\lambda \over 2 \pi}.
$$
The total density of the gas is then
$$
D= \int_{-\lambda_F}^{\lambda_F} d\mu~\rho(\mu) =
\frac{\lambda_F}{\pi}.
$$
The Fermi velocity of the gas is given by (see
Ref.~\cite{KorepinBook}) $$ v_F = \frac{D}{\pbyp{D}{\lambda_F}} =
\lambda_F,
$$
and the excitation spectrum is given by $$ \epsilon(\lambda) =
\frac{\lambda^2}{2}-h,$$ where $h$ is the chemical potential.

Thus the Bose gas with strong repulsions is equivalent to a
non-interacting Fermi gas, as is well known \cite{Girardeau1960}.

A somewhat more nontrivial regime is the one where the interaction
strength is weak compared to density, $c \ll D$. Then the Bethe
Ansatz equations can be approximately solved in the following way.

First, consider the Lieb-Liniger equation \rfs{eq:lin}, with the
kernel $K$ given by \rfs{eq:deltakernel}. It takes the form
\begin{equation} \label{eq:LLdk}
\rho(\lambda)-\frac{1}{2 \pi} \int_{-\lambda_F}^{\lambda_F} d\mu~
\frac{2 c \rho(\mu)}{(\lambda-\mu)^2+c^2}=\frac{1}{2\pi}.
\end{equation}
First we rescale variables
to find
$$
\rho(\t \lambda)-\frac{1}{2 \pi} \int_{-1}^{1} d \t \mu~ \frac{2
\t c \rho(\t \mu)}{(\t \lambda-\t \mu)^2+\t c^2}=\frac{1}{2\pi}.
$$

If $\t c$ is very small, then we use an approximation
$$
\frac{1}{2\pi} \frac{2\t c}{\t \lambda^2+\t c^2} \approx \delta(\t
\lambda)+\frac{\t c}{\pi \t \lambda^2}.
$$
This gives
$$
\int_{-1}^{1} d\t \mu~\frac{\rho(\t \mu)}{\t \lambda-\t \mu}=
\frac{\t \lambda}{2 \t c}.
$$
The solution to this well known integral equation states
\begin{equation}
\label{eq:DoSr} \rho(\t \lambda) = \frac{1}{2 \pi \t c} \sqrt{1-\t
\lambda^2},
\end{equation}
or equivalently
\begin{equation}
\label{eq:DoS} \rho(\lambda) = \frac{\lambda_F}{2 \pi c}
\sqrt{1-\frac{\lambda^2}{\lambda_F^2}}.
\end{equation}
We emphasize that \rfs{eq:DoS} is only correct for $|\lambda|
\lesssim \lambda_F$, and it definitely breaks down for $\lambda$
close to the Fermi momentum. For example, taken naively
\rfs{eq:DoS} predicts that $\rho(\lambda_F)=0$. In truth,
$\rho(\lambda_F)$ is not zero, but it is just much smaller than
$1/c$.

\rfs{eq:DoS} can be used to compute the total density
\begin{equation}\label{eq:denstotal} D= \int_{-\lambda_F}^{\lambda_F} d\mu~\rho(\mu) =
\frac{\lambda_F^2}{4 c}.\end{equation} From here we can compute the
Fermi velocity
 \begin{equation} \label{eq:vFer} v_F
= \frac{D}{\pbyp{D}{\lambda_F}} = \frac{\lambda_F}{2}.
\end{equation}
Notice that this is exactly half of the Fermi velocity in the
strong interaction case. We can now find the chemical potential
$h$ using
$$
\pbyp{h}{\lambda_F} = v_F, \rightarrow \ h=\frac{\lambda_F^2}{4}.
$$ The excitation spectrum can be found by solving the integral
equation for the excitations
\begin{equation} \label{eq:deltakernelinteg}
\epsilon(\lambda)  - \frac{1}{2 \pi} \int_{-\lambda_F}^{\lambda_F}
d\mu~ \frac{2 c
\epsilon(\mu)}{(\lambda-\mu)^2+c^2}=\frac{\lambda^2}{2}-\frac{\lambda_F^2}{4}.
\end{equation}

With the appropriate rescaling \rfs{eq:rescaling}, with the
rescaled $ \epsilon( \lambda) = q^2 \t \epsilon(\t \lambda)$, and
by taking the limit of small $c$, we bring this integral equation
to the form
$$
\int_{-1}^{1} d\t \mu~\frac{\t \epsilon(\t \mu)}{\t \lambda-\t
\mu}= \frac{1}{2 \t c} \left( \frac{\t \lambda^3}{6}- \frac{\t
\lambda}{4} \right).
$$
The solution to this equation reads \begin{equation}
\label{eq:sp1} \epsilon(\lambda) = - \frac{\lambda_F^3}{6 c}
\left(1-\frac{\lambda^2}{\lambda_F^2} \right)^{3 \over 2}, \
|\lambda| \lesssim \lambda_F.
\end{equation}
For larger values of $\lambda$ we can use the integral equation
\rfs{eq:deltakernelinteg} to find
\begin{equation} \label{eq:sp2}
\epsilon(\lambda) = \frac{\lambda_F \lambda}{2}
\sqrt{\frac{\lambda^2}{\lambda_F^2} - 1}, \ \left| \lambda \right|
\gtrsim \lambda_F.
\end{equation}
Notice that in the vicinity of $|\lambda| = \lambda_F$, both
\rfs{eq:sp1} and \rfs{eq:sp2} break down. In particular, these
expressions cannot be used to compute the Fermi velocity
\rfs{eq:vFer} simply by differentiating the energy $\epsilon$ with
respect to the momentum at $\lambda=\lambda_F$.

It is also instructive to compute the physical momentum of the
excitations $k$, using \rfs{eq:physmom}. For this purpose the
phaseshift $\theta(\lambda)$ can be replaced by $\pi \,\sign\,
\lambda$, which gives
\begin{eqnarray}
\label{eq:momnar} \t k &=&  \t \lambda + \frac{\t \lambda}{2 \t c}
\sqrt{1-\t \lambda^2} + \frac{\arcsin \t \lambda}{2 \t c}, \ |\t
\lambda| \le 1,\cr \t k &=& \t \lambda+\frac{\pi}{4 \t c}, \ |\t
\lambda| > 1.
\end{eqnarray}
Here $\t k = k/\lambda_F$.

Finally, the Luttinger parameter of the Bose gas is given by
\begin{equation} \label{eq:Le}
g=\frac{\pi D}{v_F}  \approx  \pi \sqrt{\frac{D}{ c}}.
\end{equation}
The second equality is valid only in the limit $g \gg 1$. The
Luttinger parameter at small $c$ can also be derived with the
bosonization techniques \cite{Cazalilla2004}, and the answer
coincides with \rfs{eq:Le} .

\section{Molecular Density and Three-Body Processes}

\label{AppendixF}

We would like to give an estimate of what percentage of the atomic
wave function is concentrated inside the molecules. We will do it by
directly calculating the wave function of the single-particle Fano
Anderson model \rfs{eq:Fano}. Indeed,  its Hamiltonian is quadratic,
so it can be diagonalized by the appropriate rotation of the
operators $a$, $b$. Its eigenstates at the energy $\epsilon$ can be
constructed by writing
$$\psi(\epsilon)=c_0  b^\dagger \left|0 \right\rangle +
\sum_k \alpha_k  a^\dagger_k \left| 0 \right\rangle.$$ where $c_0$
and $\alpha_k$ satisfy, curtesy of \rfs{eq:Fano}, the following
equations ($L$ is the size of the system)
\begin{eqnarray}
 \epsilon_0 ~c_0 + \frac {g}{\sqrt{L}} \sum_k \alpha_k &=& \epsilon
 ~c_0, \cr
 c_0 \frac{g}{\sqrt{L}}+\epsilon_q ~\alpha_q+\frac cL\sum_k \alpha_k
 &=& \epsilon~ \alpha_q. \nonumber
 \end{eqnarray}
It follows that
$$ \alpha_q=\frac{c}{g \sqrt{L}}
\frac{\epsilon-\epsilon_1}{\epsilon-\epsilon_q} ~c_0,
$$ while the energy level $\epsilon$ satisfies
\begin{equation} \label{eq:sm} \frac {c } L \sum_q
\frac{\epsilon-\epsilon_1}{\epsilon-\epsilon _q} =
\epsilon-\epsilon_0.
\end{equation} Here $\epsilon_1$ is given, as
before, by \rfs{eq:epsilon1}, and $\epsilon_q=q^2/(2m)$.

The normalization condition for the wave function states
$$ c_0^2 +\sum_q \alpha_q^2 = c_0^2 \left( 1+\frac{c^2}{g^2 L} \sum_q \left( \frac {\epsilon-\epsilon_1
} {\epsilon-\epsilon_q} \right)^2 \right)=1. $$ Thus the condition
that the particle spends most of its time outside the molecular
state, created by $b^\dagger$, can be formulated as follows
\begin{equation} \label{eq:tbc}
\frac{c^2}{g^2 L} \sum_q \left( \frac {\epsilon-\epsilon_1 }
{\epsilon-\epsilon_q} \right)^2 \gg 1.
\end{equation}
We would like to apply this condition to make sure that atoms spend
only a small fraction of their time bound into molecules in the
regimes of interest in this paper.

Consider first the case of small $c$. At densities up to the
critical density \rfs{eq:Dcrit0}, the particles from the Fermi sea,
having momenta from $-\lambda_F$ to $\lambda_F$, are not in
resonance with each other, so their quasistationary bound states can
be completely neglected. Suppose we add one more particle. Its
spectrum is shown on Fig.~\ref{Fig:eps2a}. If its momentum is of the
order of $2 \lambda_0$, the particle is in resonance with particles
in the Fermi sea and spends at least some of its time bound into a
molecule. This corresponds to the local minimum in
$\epsilon(\lambda)$ as shown on Fig.~\ref{Fig:eps2a}. In order to be
able to trust the physics of Fig.~\ref{Fig:eps2a}, we need to verify
that the condition \rfs{eq:tbc} holds for the majority of the two
body processes between the particle with momentum close to $2
\lambda_0$ and the particles in the Fermi sea.

We observe that at $c$ small, essentially only one term with optimal
$q$, where $\epsilon-\epsilon_q \sim c$, contributes to the sum in
\rfs{eq:sm}. We can use that to estimate $\epsilon-\epsilon_q$ and
substitute into \rfs{eq:tbc}, again retaining only one term in the
sum. We find
$$ \frac L{g^2} \left(\epsilon-\epsilon_0 \right)^2 \gg 1.
$$
This condition is clearly violated for the particles exactly at
resonance, so that $\epsilon=\epsilon_0$. Yet we only need to check
that this condition holds for the typical pair of particles, not
just those which are exactly at resonance. The typical momentum
difference for a particle at $\lambda=2\lambda_0$ and a particle in
the Fermi see is given by $2 \lambda_0 +\lambda_F$, and a typical
energy is
$$ \epsilon -\epsilon_0= \frac{\left( \lambda_0+\frac{\lambda_F}{2} \right)^2
}{2 m} -\frac{\lambda_0^2}{2m}\sim c^{\frac 13} \epsilon_0^{\frac
56} m^{\frac 16}.
$$ Here we used the estimate \rfs{eq:lcrit0} for $\lambda_F$. Hence the criterion becomes
$$ \frac L{g^2} c^{\frac23}
\epsilon_0^{\frac 53} m^{\frac 13} \gg 1.$$ Observing that $g^2 \sim
c$ (due to $\epsilon_1$, defined in \rfs{eq:epsilon1}, being finite)
we find that the above condition definitely holds at small $c$. Thus
we proved that at small $c$, in the regime of interest to us where
the density is of the order of \rfs{eq:Dcrit0}, the particle spends
the overwhelming majority of its time outside the molecular state.

The case of large $c$ is somewhat harder to analyze as we do not
have an analytic solution of this problem even with two-body
interactions neglected. Nevertheless we expect the condition
\rfs{eq:tbc} to hold as $c$ is now large, and, unless the energy
$\epsilon$ is exactly at resonance, the particle still spends most
of its time outside the molecular state.

\bibliography{bethe}

\begin{thebibliography}{16}
\bibitem[{zer()}]{zeroener}
\bibinfo{note}{A notable exception to this is given by \rfs{eq:FA} if
  $\epsilon_0-g^2/c=0$. This corresponds to the binding energy of the molecules
  being exactly zero. Under these conditions, \rfs{eq:invform} is not
  applicable and in fact, $S(0)=1$. We will not consider such situation in this
  paper, limiting ourselves to the Feshbach Hamiltonian satisfying
  \rfs{eq:nbs}}.
\end{thebibliography}

\end{document}